\documentclass[journal]{IEEEtran}
\usepackage{amsmath,amsfonts}
\usepackage{algorithmic}
\usepackage{algorithm}
\usepackage{array}
\usepackage{textcomp}
\usepackage{stfloats}
\usepackage{url}
\usepackage{verbatim}
\usepackage{graphicx}
\usepackage[sort]{cite}
\usepackage[bookmarks=false, colorlinks=false,pdfborder={0 0 0}]{hyperref} %
\usepackage{amsmath}
\usepackage{amssymb} %
\usepackage{bm}

\usepackage{balance} 
\newcommand{\ceil}[1]{\lceil {#1} \rceil}

\usepackage{lipsum} %
\usepackage[framemethod=tikz]{mdframed} %

\usepackage{pifont}%
\newcommand{\cmark}{\ding{51}}%
\newcommand{\xmark}{\ding{55}}%

\usepackage{booktabs} %
\usepackage{subcaption} %

\DeclareFontFamily{U}{mathx}{\hyphenchar\font45}
\DeclareFontShape{U}{mathx}{m}{n}{<-> mathx10}{}
\DeclareSymbolFont{mathx}{U}{mathx}{m}{n}
\DeclareMathAccent{\widebar}{0}{mathx}{"73}

\usepackage{acronym}
\acrodef{DNN}{deep neural network}
\acrodef{SNN}{slimmable neural network}
\acrodef{DSN}{dynamic slimmable network}
\acrodef{SS}{speech separation}
\acrodef{SE}{speech enhancement}
\acrodef{SI}{signal-independent}
\acrodef{SD}{signal-dependent}
\acrodef{MHA}{multi-head attention}
\acrodef{FFW}{feed-forward network}
\acrodef{MAC}{multiply-accumulate}
\acrodef{SI-SDR}{scale-invariant source-to-distortion ratio}
\acrodef{SDR}{source-to-distortion ratio}
\acrodef{TF}{time-feature}
\acrodef{SIR}{signal-to-interference ratio}
\acrodef{STFT}{short-time Fourier transform}
\acrodef{GLN}{global layer normalization}
\acrodef{DOA}{direction of arrival}
\acrodef{SC}{single-channel}
\acrodef{MC}{multi-channel}
\acrodef{BF}{beamformer}
\acrodef{STFT}{short-time Fourier transform}
\acrodef{PIT}{permutation invariant training}

\begin{document}
\title{Dynamic Slimmable Networks for \\ Efficient Speech Separation} %

\author{Mohamed~Elminshawi,~\IEEEmembership{Student Member,~IEEE,}
        Srikanth~Raj~Chetupalli,~\IEEEmembership{Member,~IEEE,}
        and~Emanu\"{e}l~A.~P.~Habets,~\IEEEmembership{Senior Member,~IEEE}%
\thanks{Mohamed~Elminshawi and Emanu\"{e}l~A.~P.~Habets are with the International Audio Laboratories Erlangen (a joint institution of the Friedrich-Alexander-Universität Erlangen-Nürnberg (FAU) and Fraunhofer IIS), 91058 Erlangen, Germany (\mbox{e-mail}: \href{mailto:mohamed.elminshawi@audiolabs-erlangen.de}{\nolinkurl{mohamed.elminshawi@audiolabs-erlangen.de}}, 
\href{mailto:emanuel.habets@audiolabs-erlangen.de}{\nolinkurl{emanuel.habets@audiolabs-erlangen.de}}). 
Srikanth~Raj~Chetupalli is with the Indian Institute of Technology Bombay, Mumbai, India (\mbox{e-mail}: \href{mailto:srikanth.chetupalli@iitb.ac.in}{\nolinkurl{srikanthrajch@iitb.ac.in}}). This work was conducted while he was with the International Audio Laboratories Erlangen.}%
}


%
%
%

\maketitle

\begin{abstract}

Recent progress in speech separation has been largely driven by advances in deep neural networks, yet their high computational and memory requirements hinder deployment on resource-constrained devices. A significant inefficiency in conventional systems arises from using static network architectures that maintain constant computational complexity across all input segments, regardless of their characteristics. This approach is sub-optimal for simpler segments that do not require intensive processing, such as silence or non-overlapping speech. To address this limitation, we propose a dynamic slimmable network (DSN) for speech separation that adaptively adjusts its computational complexity based on the input signal. The DSN combines a slimmable network, which can operate at different network widths, with a lightweight gating module that dynamically determines the required width by analyzing the local input characteristics. To balance performance and efficiency, we introduce a signal-dependent complexity loss that penalizes unnecessary computation based on segmental reconstruction error. Experiments on clean and noisy two-speaker mixtures from the WSJ0-2mix and WHAM! datasets show that the DSN achieves a better performance-efficiency trade-off than individually trained static networks of different sizes.

\end{abstract}


\section{Introduction}
\IEEEPARstart{D}{eep} neural networks (DNNs)\acused{DNN} have become ubiquitous in modern speech processing applications due to their powerful modeling capabilities, driving significant progress in tasks such as \acl{SE}~\cite{xu2014regression, zheng2023sixty}, automatic speech recognition~\cite{nassif2019speech, prabhavalkar2023end}, and keyword spotting~\cite{lopez2021deep}. In particular, \ac{SS} — the task of isolating individual speakers from a mixture — has seen remarkable improvements due to advances in \acp{DNN}~\cite{wang2018supervised, Luo2019a, subakan2021attention, subakan2023exploring, wang2023tf, chetupalli2023speaker}. However, deploying \acp{DNN} on resource-constrained devices (e.g., smartphones and hearing aids) remains challenging due to limited computational and memory resources. Moreover, conventional \ac{DNN}-based speech processing systems typically employ static networks during inference, resulting in a constant computational cost across all segments of the input signal. This uniformity is computationally inefficient, particularly for segments requiring less intensive processing, such as those containing silence or clean speech.

In contrast to traditional static networks, dynamic networks are designed to adapt their computational graph during inference in an input-dependent fashion~\cite{han2021dynamic}. This adaptability allows the computational cost to vary based on the characteristics of each input, thereby improving the computational efficiency. Dynamic networks can generally achieve adaptability through three main mechanisms: dynamic routing, dynamic depth, and dynamic width, each controlling a different aspect of the network's structure at inference time~\cite{han2021dynamic}. 

Dynamic routing selects different computational paths within a network based on the input. A key example is the mixture-of-experts framework~\cite{jacobs1991adaptive, fedus2022switch, you2021speechmoe}, where inputs are routed to different sub-networks (i.e., experts), enabling each expert to specialize in different parts of the input space. In contrast, dynamic depth approaches, such as early exiting~\cite{bolukbasi17a} and layer skipping~\cite{Wang_2018_ECCV, campos2017skip}, adapt the number of executed layers according to the input characteristics, thereby reducing unnecessary computations and improving efficiency. Finally, dynamic width enables networks to select a subset of layer parameters (e.g., neurons or channels) based on the input at inference time. A notable example is \acp{DSN}~\cite{li2021dynamic}, where the network width is dynamically adjusted by slicing layer parameters. Unlike other dynamic width approaches~\cite{chen2019you, hua2019channel}, the slimming technique is particularly advantageous for hardware implementation, as it simplifies memory access without the need for indexing~\cite{li2021dynamic}.

In this paper, we introduce a \ac{DSN} for \ac{SS} that adaptively adjusts its computational complexity at inference time based on the input signal. Unlike static networks with fixed computational complexity, the proposed \ac{DSN} adapts its network width on a per-frame basis, enabling it to allocate more resources to challenging segments and reduce computation on simpler ones. This adaptive behavior is driven by a lightweight gating module that analyzes local characteristics of the input signal and predicts the appropriate width for each time frame. To further promote efficiency, we propose a \ac{SD} complexity loss that discourages unnecessary computations in segments where the network already performs well, as measured by the segmental reconstruction error. Experimental evaluation on clean and noisy two-speaker mixtures from the WSJ0-2mix~\cite{hershey2016deep} and WHAM!~\cite{wichern2019wham} datasets demonstrates that the proposed \ac{DSN} achieves a better trade-off between separation performance and computational efficiency than individually trained static networks of different sizes.

The remainder of this paper is organized as follows. Section~\ref{sec:related_work} provides a brief review of related work. In Section~\ref{sec:proposed_method}, we formally define the \ac{SS} problem and present the proposed \ac{DSN}. Section~\ref{sec:exp_setup} describes the experimental setup. In Section~\ref{sec:exp_results}, we present and analyze the experimental results. Finally, the findings are discussed in Section~\ref{sec:discussion}.

\section{Related Work}
\label{sec:related_work}
\subsection{Dynamic Inference for Speech Processing}
In the context of monaural \ac{SS}, several studies have investigated the use of dynamic inference to improve the computational efficiency of \acp{DNN}. For instance, early exiting has been employed in~\cite{chen2021don, bralios2022latent}, to reduce computational cost by terminating execution at an early stage, either when two consecutive layers produce sufficiently similar 
predictions~\cite{chen2021don}, or by using a gating module that predicts whether to stop execution at certain layers~\cite{bralios2022latent}. However, both studies utilized dynamic inference at the utterance level, which limits their ability to achieve higher computational efficiency by adapting to local signal characteristics. Frame-level adaptation has been explored in a few studies in the context of \acl{SE}, such as skipping hidden state updates in recurrent networks~\cite{le2022inference}, or deactivating unnecessary convolutional channels via dynamic channel pruning~\cite{miccini2024scalable}.

\subsection{Relation to our Previous Studies}
This study builds on our earlier research on \acp{SNN} and \acp{DSN} for \ac{SS}. In a prior study~\cite{elminshawi2023slimtasnet}, we presented a \ac{SNN} that can operate at different widths of its layers to enable a flexible performance-efficiency trade-off during inference. However, width adaptation was performed at the utterance level and was determined by an external factor (e.g., resource budget), rather than based on the characteristics of the input signal.

More recently, we introduced a \ac{DSN} for \ac{SS}~\cite{elminshawi2024dynamic}, where the network width was adapted based on local input characteristics. Specifically, dynamic inference was demonstrated using the SepFormer architecture~\cite{subakan2021attention}, which consists of transformer layers applied within a dual-path framework. The complexity of the intra-chunk transformer layers was controlled by adapting the number of neurons in the intermediate layer of the \ac{FFW} blocks using a block-specific gating module that analyzes each input chunk. 

The current study extends~\cite{elminshawi2024dynamic} by incorporating several key architectural and methodological advancements.
In particular, instead of using a block-specific gating module, we employ a single gating module that analyzes the local characteristics of the input mixture and predicts the required width for each time frame. Additionally, width adaptation is applied not only to the \ac{FFW} but also to the \ac{MHA} block, and is further extended to the inter-chunk transformer layers. These design choices allow for a greater reduction in computational cost.

On the training side, we introduce a novel \acl{SD} complexity loss, which adjusts the regularization strength based on the segmental reconstruction quality as a proxy for the difficulty of each input segment. This encourages the \ac{DSN} to allocate more computation to challenging segments and less computation to simpler ones. 
We conduct an extensive ablation study to better understand different aspects of the proposed \ac{DSN}, including the impact of the complexity loss. Evaluation is also extended to include mixtures with various overlap ratios and additive background noise from the WHAM! dataset~\cite{wichern2019wham}.

\section{Proposed Method}
\label{sec:proposed_method}
We begin by formally defining the \ac{SS} problem, followed by a detailed description of the proposed \ac{DSN}.

\subsection{Neural Speech Separation}
We consider a single-channel signal $y \in \mathbb{R}^L$ of length $L$ samples containing overlapping speech from $J$ speakers and additive background noise, i.e.,
\begin{equation}
    y = \sum_{j=1}^J s_j + v,
\end{equation}
where $s_j \in \mathbb{R}^L$ denotes the speech component of the \mbox{$j$-th} speaker, and $v \in \mathbb{R}^L$ is the noise component. A \ac{SS} system, represented by $\mathcal{F}$, aims to reconstruct the speech signals of all speakers from the observed mixture signal $y$ as
\begin{equation}
    \{ \widehat{s}_j \}_{j=1}^{J} = \mathcal{F} \left(y \right),
\end{equation}
where $\widehat{s}_j \in \mathbb{R}^L$ denotes the estimated signal of the $j$-th speaker.  In this study, we consider a data-driven approach to \ac{SS} using a \ac{DNN} for the system $\mathcal{F}$. To demonstrate the utility of dynamic inference for \ac{SS}, we adopt SepFormer~\cite{subakan2021attention} as a representative \ac{DNN} architecture and modify its transformer layers to realize a \ac{DSN}. However, the proposed approach can also be applied to other architectures, for example, by selectively utilizing a subset of channels in convolutional layers, as demonstrated in~\cite{elminshawi2023slimtasnet, miccini2024scalable}.

\subsection{System Overview}
\begin{figure*}[t]
    \centering
    \subfloat[DSN\label{fig:case_a}]{
        \includegraphics[height=7cm]{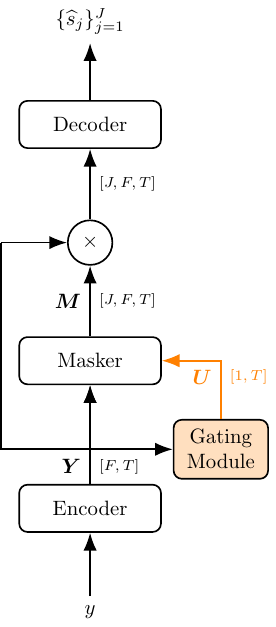}}
    \hfill
    \subfloat[Gating Module\label{fig:case_b}]{
        \includegraphics[height=7cm]{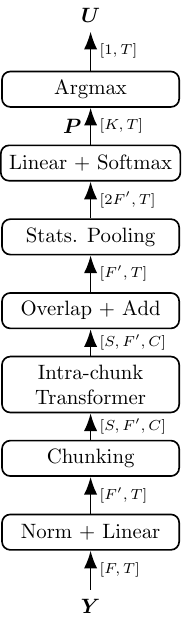}}
    \hfill
    \subfloat[Masker\label{fig:case_c}]{
        \includegraphics[height=7cm]{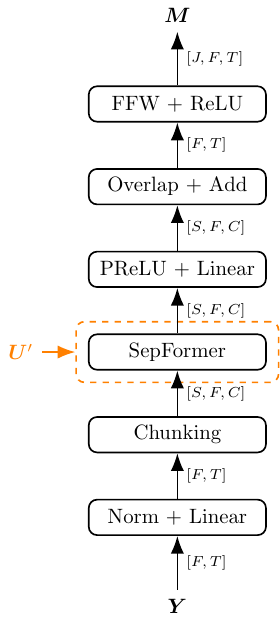}}
    \hfill
    \subfloat[SepFormer\label{fig:case_d}]{
        \includegraphics[height=7cm]{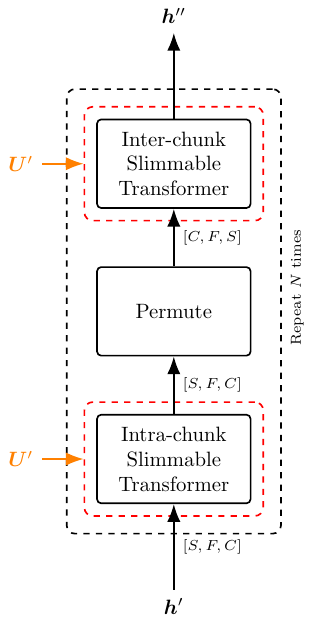}}
    \hfill
    \subfloat[Slimmable Transformer\label{fig:case_e}]{
        \includegraphics[height=7cm]{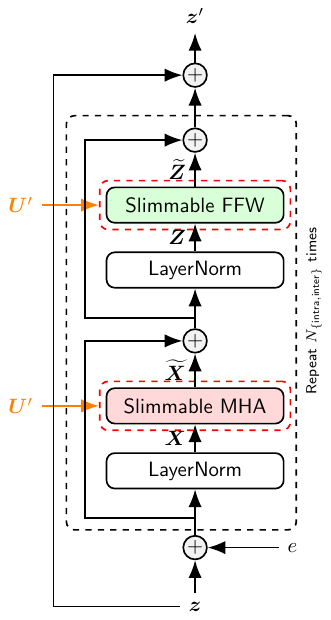}}
    \caption{Overview of the proposed \ac{DSN} for \ac{SS}. (a) The \ac{DSN} consists of an encoder-masker-decoder network and a lightweight gating module that adapts the computational complexity of the masker. (b) The gating module analyzes the local characteristics of the encoded mixture $\bm{Y}$ and predicts a utilization sequence $\bm{U} = [U_1, U_2, \ldots, U_T] \in \mathbb{R}^{1 \times T}$, where each utilization factor $U_t \in (0,1]$ corresponds to time frame $t$. The masker in (c) is based on the SepFormer architecture~\cite{subakan2021attention} that operates within a dual-path structure~\cite{Luo2019a} and is composed of slimmable intra- and inter-chunk transformer blocks (shown in d), to model short- and long-term characteristics, respectively. Due to the dual-path structure, the chunking operation is also applied to predicted utilization $\bm{U}$, resulting in a chunked utilization tensor $\bm{U}' \in \mathbb{R}^{1\times S \times C}$. (e) The chunked utilization $\bm{U}'$ modifies the computational complexity of the slimmable \ac{MHA} and \ac{FFW} blocks in the slimmable transformer layers.}
    \label{fig:block_diagram}
\end{figure*}

Figure~\ref{fig:block_diagram} shows an overview of the proposed \ac{DSN} for \ac{SS}. It follows an encoder-masker-decoder structure (Figure~\ref{fig:block_diagram}a), where the encoder and decoder operate in the time domain using convolutional layers trained jointly with the masker in an end-to-end fashion. Given the input mixture $y$, the encoder produces a \ac{TF} representation $\bm{Y}  \in \mathbb{R}^{F\times T}$, where $F$ is the number of features and $T$ represents the number of time frames. The masker generates a stack of \ac{TF} masks, one for each speaker, forming a tensor $\bm{M} \in \mathbb{R}^{J\times F \times T}$.
Each mask $\bm{M}_j \in \mathbb{R}^{F \times T}$ is multiplied element-wise with $\bm{Y}$ and fed to the decoder to reconstruct the corresponding time-domain signal $\widehat{s}_j$.

The masker module, illustrated in Figure~\ref{fig:block_diagram}c, is based on SepFormer~\cite{subakan2021attention}, which adopts the dual-path framework~\cite{Luo2019a} to capture both short- and long-term dependencies via intra- and inter-chunk blocks (Figure~\ref{fig:block_diagram}d). In SepFormer~\cite{subakan2021attention}, these blocks contain transformer layers with a fixed structure during training and inference, and thus we refer to them as static. In this work, we replace the static transformer layers in the masker with slimmable transformer layers (Figure~\ref{fig:block_diagram}e). In particular, both the \ac{MHA} and \ac{FFW} blocks are made slimmable, allowing the computational complexity to vary dynamically during inference.

To control the computational complexity of the masker in an input-dependent fashion, a gating module (Figure~\ref{fig:block_diagram}b) analyzes the short-term characteristics of the encoded mixture $\bm{Y}$ and predicts a utilization sequence $\bm{U} = [U_1, U_2, \ldots, U_T] \in \mathbb{R}^{1 \times T}$, where each utilization factor $U_t \in (0,1]$ corresponds to time frame $t$. Since the transformer blocks in SepFormer operate within a dual-path framework that involves intra- and inter-chunk processing, the predicted utilization sequence $\bm{U}$ is also segmented using a chunking operation similar to the input features in the masker. This yields a chunked utilization tensor $\bm{U}' \in \mathbb{R}^{1\times S \times C}$, where $C$ denotes the chunk size, and $S$ represents the number of chunks. The chunked utilization $\bm{U}'$ is then used to adapt the width of the slimmable transformer layers in both intra- and inter-chunk paths.

To demonstrate how the utilization factor affects the computational and memory complexity of the \ac{DSN}, Table~\ref{tab:complexity_params_sepformer} shows the number of \ac{MAC} operations per second and the number of active parameters at different utilization levels. As shown, decreasing the utilization factor proportionately reduces the computational complexity and the effective model size.

In the following, we first discuss the gating module, review the static transformer layer in SepFormer, and then present our modifications to transform it into a slimmable layer.

\begin{table}[t]
    \centering
    \setlength{\tabcolsep}{10pt}
    \renewcommand{\arraystretch}{1.5}
    \caption{Computational and memory complexity of the \ac{DSN} at various fixed utilization factors. GMAC/s is computed for a four-second input signal. Params refers to the number of active parameters.}
    \begin{tabular}{c c c}
        \toprule
        \textbf{Utilization} & \textbf{GMAC/s} & \textbf{Params (M)} \\ \midrule
        0.125 & \phantom{0}5.8 & \phantom{0}2.0 \\
        0.25\phantom{0}  & \phantom{0}9.0 & \phantom{0}3.6 \\
        0.5\phantom{00}   & 15.4 & \phantom{0}6.7 \\
        0.75\phantom{0}  & 21.8 & \phantom{0}9.9 \\
        1.0\phantom{00}   & 28.1 & 13.0 \\ \bottomrule
    \end{tabular}
    \label{tab:complexity_params_sepformer}
\end{table}

\subsection{Gating Module}
We employ a single gating module, denoted by $\mathcal{G}$, to control the computational complexity of the slimmable layers in the \ac{DSN}.  The gating process is formulated as a classification task over a discrete set of predefined utilization factors, denoted by $\mathcal{U}=\{u_k\}_{k=1}^K$, where $K$ denotes the number of possible utilization levels. As shown in Figure~\ref{fig:block_diagram}b, the gating module takes as input the mixture representation $\mathbf{Y}$ and estimates a probability vector for each time frame $t\in \{1,2, \ldots,T\}$, i.e., $\bm{P} = [P_1, \ldots, P_T] \in \mathbb{R}^{K\times T}$, as
\begin{equation}
    \bm{P} = \mathcal{G}(\bm{Y}),
\end{equation}
\noindent
where $P_{t,k}$ denotes the probability assigned to utilization level $u_k$ at time frame $t$.
The output utilization factor at time frame $t$, i.e., $U_t$, is then obtained by selecting the utilization level corresponding to the maximum probability, i.e., 
\begin{align}
    k^*_t &= \operatorname*{argmax}_k P_{t,k}, \\
    U_t &= u_{k^*_t}.
    \label{eq:internal_gating}
\end{align}

During training, we apply the Gumbel softmax trick~\cite{jang2017categorical}, which approximates sampling from a categorical distribution in a differentiable manner. Specifically, given the logits $R_{t,k}$ produced by the final linear layer of the gating module for each utilization level $u_k$ at time frame $t$, the soft probability vector $P_t \in \mathbb{R}^K$ is computed as
\begin{equation}
    P_{t,k} = \frac{e^{\left(R_{t,k} + g_k\right)/\tau}}{\sum_{k'=1}^K e^{\left(R_{t,k'} + g_{k'}\right)/\tau}},
\end{equation}
\noindent
with $g_k \sim \text{Gumbel}(0,1)$ denoting a Gumbel noise sample, and $\tau \geq 1$ being the temperature parameter. To obtain discrete decisions while preserving gradient flow, the straight-through estimator (STE)~\cite{bengio2013estimating} is used. In the forward pass, a hard one-hot vector is obtained via the argmax operation, while in the backward pass, gradients are propagated through the soft probabilities. During inference, no Gumbel noise is sampled, and the gating module behaves deterministically. 

The structure of the gating module (Figure~\ref{fig:block_diagram}b) resembles that of the masker, but employs only a lightweight intra-chunk transformer. In addition, a statistics pooling layer is applied to aggregate the mean and standard deviation of the features within a short context window, followed by a linear classification layer. Further details about the configuration of the gating module are provided in Section~\ref{sec:model_cfgs}.

\subsection{Static Transformer Layer}
\label{sec:static_transformer}
The core components of a transformer layer~\cite{vaswani2017attention} are the \ac{MHA} and the position-wise \ac{FFW}, supplemented by positional encoding, layer normalization, and residual connections. 
Let $\bm{X} = [X_1, \dots, X_M] \in \mathbb{R}^{F\times M}$ denote the input to the \ac{MHA} block, where $F$ denotes the feature dimension and $M$ represents the sequence length. Note that the sequence length corresponds to the chunk size (i.e., $M=C$) for intra-chunk transformers and to the number of chunks (i.e., $M=S$) for inter-chunk transformers. 
The \ac{MHA} comprises $H$ heads, each consisting of three linear layers, parameterized by the weight tensors $W^h_q, W^h_k, W^h_v \in \mathbb{R}^{d\times F}$, for the queries, keys, and values, respectively. Here, $d$ represents the embedding dimension for each head, and $h$ denotes the head index. For each head $h \in \{1,2,\dots,H\}$ and each frame $m \in \{1,2,\dots,M\}$, the queries ($Q^h_m \in \mathbb{R}^d$), keys ($K^h_m \in \mathbb{R}^d$), and values ($V^h_m \in \mathbb{R}^d$) are computed as
\begin{equation}
    Q^h_m = W^h_q X_m, \quad K^h_m = W^h_k X_m, \quad V^h_m = W^h_v X_m.
    \label{eq:qkv}
\end{equation}
\noindent
Self-attention scores at frame $m$ are computed using the scaled dot-product between the query $Q_m^h$ and all keys $K_j^h$, for $j \in \{1,2,\dots,M\}$, as
\begin{equation}
    \alpha_{m,j}^h = \frac{{Q_m^h}^\mathrm{T} K_j^h}{\sqrt{d}},
\end{equation}
where $(\cdot)^\mathrm{T}$ denotes the transpose operation.
\noindent
The attention scores are then normalized using the softmax function across all frames, i.e.,

\begin{equation}
    A_{m,j}^h = \frac{e^{\alpha_{m,j}^h}}{\sum_{j'=1}^M e^{\alpha_{m,j'}^h}}.
    \label{eq:mha_softmax}
\end{equation}
\noindent
The output $\widebar{V}_m^h$ at frame $m$ is obtained by aggregating the value vectors weighted by their attention scores, i.e.,
\begin{equation}
    \widebar{V}_m^h = \sum_{j=1}^{M} A^h_{m,j}\,V^h_{j}.
    \label{eq:mha_active}
\end{equation}
\noindent
Subsequently, the outputs from all attention heads $\widebar{V}_m^h$ are combined through head-specific linear layers, parametrized by $W_o^h \in \mathbb{R}^{F\times d}$, and aggregated as
\begin{equation}
    \widetilde{X}_m = \sum_{h=1}^H W_o^h\,\widebar{V}_m^h.
    \label{eq:summation_form}
\end{equation}
\noindent
Note that the summation form in (\ref{eq:summation_form}) is mathematically equivalent to the standard concatenation form~\cite{jin2024moh}.

The second key component of the transformer layer is the position-wise \ac{FFW}, which consists of two linear layers with a ReLU activation in between. These layers are parametrized by $W_1 \in \mathbb{R}^{D\times F}$ and $b_1 \in \mathbb{R}^{D}$ for the first layer, and $W_2 \in \mathbb{R}^{F\times D}$ and $b_2 \in \mathbb{R}^{F}$ for the second, where $D$ denotes the dimensionality of the intermediate layer. Given an input $Z_m \in \mathbb{R}^{F}$ to the \ac{FFW} at frame $m$, the output $\widetilde{Z}_m \in \mathbb{R}^{F}$ is computed as
\begin{equation}
    \widetilde{Z}_m = W_2 \,\text{ReLU}(W_1\,Z_m + b_1) + b_2.
\end{equation}

\subsection{Slimmable Transformer Layer}
\label{sec:slim_transformer}
The static transformer layer is replaced with a slimmable variant whose computational complexity is adaptively controlled by the gating module. As described earlier, the predicted utilization sequence $\bm{U} \in \mathbb{R}^{1 \times T}$ is segmented into a chunked tensor \( \bm{U}' \in \mathbb{R}^{1 \times S \times C} \) to align with the dual-path processing structure. We denote by $U'_m$, for $m \in \{1, 2, \ldots, M\}$, the utilization factor at the $m$-th frame within a chunk, where $M=C$ for intra-chunk blocks and $M=S$ for inter-chunk blocks. The following sections describe how the utilization factor $U'_m$ adapts the slimmable \ac{MHA} and \ac{FFW} blocks.

\subsubsection{Slimmable Multi-head Attention (MHA)}
\begin{figure}[t]
\centering
  \includegraphics[width=0.8\linewidth]{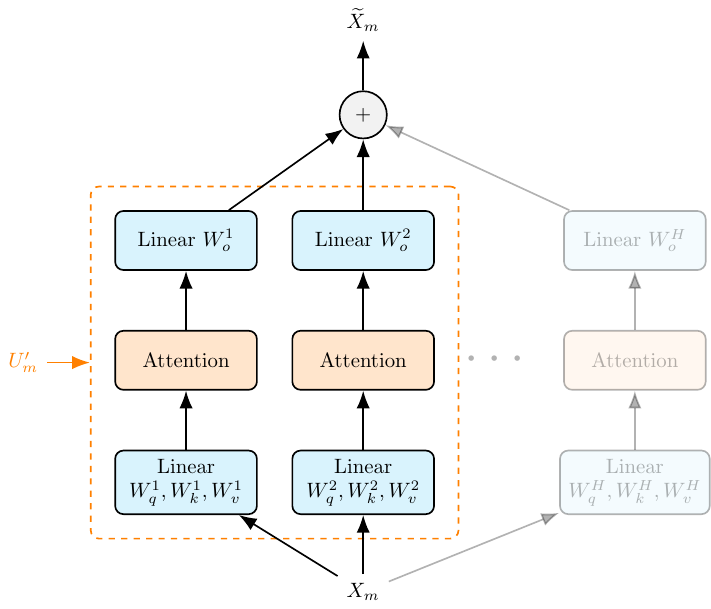} 
  \caption{Illustration of the slimmable \acf{MHA} block. The utilization factor $U'_m$ determines the number of attention heads used to process each input $X_m$. Only the selected heads (enclosed in the orange dashed box) are executed, while the remaining heads are skipped, enabling dynamic adaptation of computational cost. Self-attention is computed only on frames where the corresponding head is active, i.e., $m \in \mathcal{T}^h$, as described in (\ref{eq:selfattention_active}).}
  \label{fig:slimmable_MHA}
\end{figure}

The static \ac{MHA} block is made slimmable by controlling the number of attention heads, activated based on the utilization factor $U'_m$ for each input $X_m$, as illustrated in Figure~\ref{fig:slimmable_MHA}. In particular, only the first $\ceil{U'_m H}$ heads are utilized, where $\ceil{\cdot}$ denotes the ceiling operator. To account for the frame-wise head activation, we define for each head $h$ the set of frames where it is active as $\mathcal{T}^h=\{m \in \{1, 2, \dots, M\}\,|\,h \leq \ceil{U'_m H} \}$. This set determines the frames over which each head performs input projections, attention score normalization, and value aggregation. Specifically, the input projections in (\ref{eq:qkv}) are computed only at frames $m \in \mathcal{T}^h$.  Similarly, the softmax normalization in~(\ref{eq:mha_softmax}) is restricted to the frames $m \in \mathcal{T}^h$, i.e., 
\begin{equation}
    A_{m,j}^h = \frac{e^{\alpha_{m,j}^h}}{\sum_{j' \in \mathcal{T}^h} e^{\alpha_{m,j'}^h}}.
    \label{eq:selfattention_active}
\end{equation}
\noindent
The output $\widebar{V}_m^h$ at frame $m \in \mathcal{T}^h$ is computed as
\begin{equation}
    \widebar{V}_m^h = \sum_{j \in \mathcal{T}^h} A^h_{m,j}\,V^h_{j}.
\end{equation}

\noindent
Finally, the output projection in (\ref{eq:summation_form}) is computed using only the active heads at frame $m$, i.e., 
\begin{equation}
    \widetilde{X}_m = \sum_{h=1}^{\ceil{U'_m H}} W_o^h\,\widebar{V}_m^h.
    \label{eq:slim_heads}
\end{equation}
\noindent

\subsubsection{Slimmable Position-wise Feed-forward Network (FFW)}
\begin{figure}[t]
\centering
  \includegraphics[width=\linewidth]{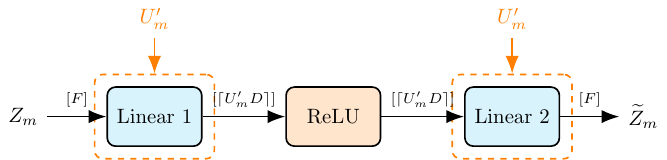} 
  \vspace{0.1em}
  \caption{Structure of the slimmable \acf{FFW}. For each input $Z_m$, the number of neurons in the intermediate layer is dynamically determined by the utilization factor $U'_m$, such that only the first $\ceil{U'_m D}$ units are used. 
  }
  \label{fig:slimmable_FFW}
\end{figure}

In the slimmable \ac{FFW}, the utilization factor $U'_m$ determines the number of neurons in the intermediate layer used to process the input $Z_m$, as shown in Figure~\ref{fig:slimmable_FFW}.
Specifically, only the first $\ceil{U'_m D}$ neurons are utilized in computing the layer output. 
This is realized by processing the input $Z_m$ using a subset of the parameters $W_1$, $b_1$, and $W_2$, denoted by $\widetilde{W}_1 \in \mathbb{R}^{\ceil{U'_m D}\times F}$, $\widetilde{b}_1 \in \mathbb{R}^{\ceil{U'_m D}}$, and $\widetilde{W}_2 \in \mathbb{R}^{F\times \ceil{U'_m D}}$, respectively. The output at frame $m$ is computed as
\begin{equation}
    \widetilde{Z}_m = \widetilde{W}_2\,\text{ReLU}(\widetilde{W}_1\,Z_m + \widetilde{b}_1) + b_2.
\end{equation}

\subsection{Optimization}
The proposed \ac{DSN} is optimized using two loss functions to balance separation performance and computational efficiency. The primary loss is the signal reconstruction loss, which is based on the \ac{SI-SDR}~\cite{leroux2019} and computed in a permutation-invariant manner at the utterance level~\cite{kolbaek2017multitalker}, i.e.,
\begin{equation}
    \mathcal{L}_{\text{signal}} = \min_{q \in \mathcal{Q}} \frac{1}{J} \sum_{j=1}^J -\text{SI-SDR}(\widehat{s}_j, s_{q(j)}),
\end{equation}
\noindent
where $q:\{1, \ldots, J\} \rightarrow \{1, \ldots, J\}$ is a bijective mapping between the speakers and estimated signals, and $\mathcal{Q}$ denotes the set of all possible mappings (i.e., permutations).

To encourage the gating module to predict lower utilization factors and thereby improve computational efficiency, we introduce a computational complexity loss, defined as
\begin{equation}
\label{eq:complexity}
    \mathcal{L}_{\text{complexity}} = \frac{1}{T} \sum_{t=1}^T w_{t} \left(U_t - \min(\mathcal{U}) \right)^2,
\end{equation}

\noindent
where $U_t$ denotes the predicted utilization at time frame $t$, and $\min(\mathcal{U})$ represents the minimum possible utilization in the set $\mathcal{U}$. 
For the weighting factor $w_{t}\geq0$, we consider two types of weighting schemes: a \ac{SI} case, referred to as $w^{\text{SI}}$, where a constant weight of $1$ is applied across all time frames (i.e., \( w_t = 1 \, \forall t \)), and a \ac{SD} case, denoted by $w_t^{\text{SD}}$, where the weight varies at each time frame based on the segmental reconstruction quality, as described in Section~\ref{sec:sd_wf}. The minimum of $\mathcal{L}_{\text{complexity}}$ is achieved when the minimum possible utilization is predicted for all time frames. 

The overall training objective is defined as a weighted combination of the signal reconstruction and complexity losses, i.e., 
\begin{equation}
\label{eq:overall_loss}
    \mathcal{L} = \mathcal{L}_{\text{signal}} +  \alpha\,\mathcal{L}_{\text{complexity}},
\end{equation}

\noindent
where $\alpha$ is a hyperparameter that controls the contribution of the complexity loss to the total loss. 

\subsection{Signal-dependent Weighting Factor}
\label{sec:sd_wf}
\begin{figure}[t]
\centering
  \includegraphics[width=\linewidth]{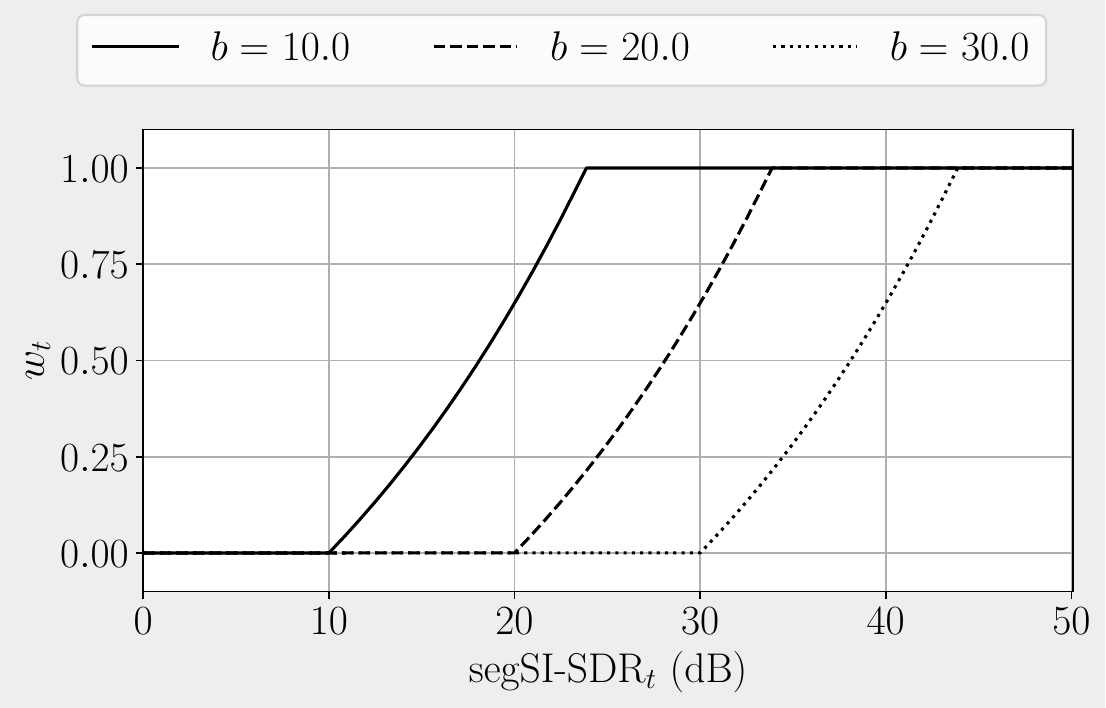} 
  \caption{\Acf{SD} weighting factor $w_t$ as a function of segmental SI-SDR at time frame $t$ (segSI-SDR$_t$), as defined in (\ref{eq:sd_wf_eq}), for different values of $b$. The curves are plotted for $a=0.05$.
  }
  \label{fig:dynamic_scaling}
\end{figure}

For the weighting factor $w_t$ in (\ref{eq:complexity}), we propose a \ac{SD} formulation based on the segmental quality of the reconstructed signals, measured by segmental \ac{SI-SDR} (segSI-SDR). This weighting scheme encourages lower utilization when the separation quality is already high, thereby saving computational resources without sacrificing performance. Specifically, we define the weighting factor as
\begin{equation}
    w_t = \begin{cases} 
      0 &  \text{segSI-SDR}_t < b, \\
      \min\bigl( e^{a\,(\text{segSI-SDR}_t - b) - 1},\; 1 \bigr) & \text{segSI-SDR}_t \geq b,
   \end{cases}
   \label{eq:sd_wf_eq}
\end{equation}
\noindent
where segSI-SDR$_t$ denotes the segmental SI-SDR at time frame $t$. The parameter $a$ controls the steepness of the exponential curve, while $b$ defines the threshold after which the weighting factor is non-zero. Figure~\ref{fig:dynamic_scaling} shows $w_t$ for different values of $b$.

The segSI-SDR$_t$ is computed as the average segmental SI-SDR between each estimated signal and its corresponding ground-truth signal. For segments with a single active speaker, the segSI-SDR score of the corresponding estimated signal is used. For silence segments, the weighting factor is set to its maximum value, i.e., $w_t=1$. 

\section{Experimental Setup}
\label{sec:exp_setup}

\subsection{Datasets}
Speech mixtures from the WSJ0-2mix dataset~\cite{hershey2016deep}  were used in our experiments. The official splits consist of $20$k, $5$k, and $3$k two-speaker mixtures (i.e., $J=2$) for training, validation, and testing, respectively. The signal-to-interference ratio ranges from $0$~dB to $5$~dB. The sampling frequency of the audio files is $8$~kHz. We used the \texttt{min} version which corresponds to fully overlapping mixtures (i.e., $100\%$~overlap). To evaluate the proposed method across a broader range of overlap scenarios, we additionally generated partially overlapping mixtures by temporally offsetting the start time of one speaker’s signal relative to that of the other. The offset was computed to achieve a specific target overlap ratio, defined as the ratio of the speech overlap duration (i.e., when both speakers are simultaneously active) to the effective duration of the shorter utterance (i.e., after excluding leading and trailing silences). A voice activity detector\footnote{\url{https://github.com/snakers4/silero-vad}} was used to exclude these silences when estimating the effective speech duration, which in turn determined the offset required to achieve the desired overlap.

We additionally experimented with more realistic speech mixtures using the WHAM! dataset~\cite{wichern2019wham}, which combines the two-speaker mixtures from WSJ0-2mix with additive background noise recorded in urban environments (e.g., coffee shops and restaurants). In the official WHAM! dataset, the background noise spans the entire duration of each mixture. To assess performance under more varied conditions, we generated mixtures with reduced noise activity, in which the noise is present only during a fraction of the total mixture duration.

\subsection{Implementation Details}
\label{sec:model_cfgs}
For the SepFormer architecture~\cite{subakan2021attention}, we used the implementation provided by the SpeechBrain toolkit~\cite{SB2021}. Following the notations of SepFormer~\cite{subakan2021attention}, we set the hyperparameters as follows: the number of repetition blocks ($N=2$), number of layers ($N_\text{intra}=N_\text{inter}=4$), number of heads ($H_\text{intra}=H_\text{inter}=8$), dimensionality of the intermediate layer of the \ac{FFW} ($D_\text{intra}=D_\text{inter}=1024$), and chunk size~$C=50$~($51$~ms). The encoder and decoder are symmetric, with the window size, hop size, and number of channels ($F$) set to~$16$~($2$~ms), $8$~($1$~ms), and $256$, respectively.

For the gating module, the hyperparameters were set as: $F'=32$, $N_\text{intra}=1$, $H_\text{intra}=1$, $D_\text{intra}=128$, and $\tau=1$. The gating module contained approximately $21.7$k parameters. The size of the context window for the statistics pooling layer was $64$ frames ($65$~ms). The segSI-SDR was computed on $96$~ms segments with $50\%$ overlap. 

\subsection{Training Setup}
\label{sec:training_setup}
To optimize the different models, we used the Adam optimizer~\cite{Kingma2015}. The initial learning rate was set to~$1.5\times 10^{-4}$ and reduced by a factor of~$2$ if the validation loss did not decrease for~$6$ consecutive epochs. The maximum number of epochs was set to~$200$. A batch size of~$2$~samples was used, each with a duration of~$4$~s. Early stopping was applied with a patience of~$30$~epochs. Gradients were clipped if their \mbox{$\ell_2$-norm} exceeded a value of~$5$. For each mixture, the overlap ratio was uniformly sampled from $[0.25, 1.0]$. For WHAM! mixtures~\cite{wichern2019wham}, the noise activity level was sampled from $[0.0, 1.0]$.

During training, we applied \emph{gating dropout} by randomly ignoring the predicted gates with a probability of $30\%$. In such cases, a utilization level was uniformly sampled from the set $\mathcal{U}$ and applied across all time frames. This strategy allows the \ac{DSN} to generalize to externally provided gating decisions during inference, as discussed in Section~\ref{sec:robustness}.

\section{Experimental Results}
\label{sec:exp_results}
We evaluated the separation performance of the proposed \ac{DSN} using the \ac{SI-SDR}~\cite{leroux2019} to assess the quality of the reconstructed speech signals. The computational complexity was measured in terms of the number of \ac{MAC} operations per second, computed using the \texttt{torchprofile} package\footnote{\url{https://github.com/zhijian-liu/torchprofile}}. Audio examples are available online\footnote{\url{https://www.audiolabs-erlangen.de/resources/2025-TASLP-DSN}}.

\subsection{Impact of Complexity Loss}
\begin{figure}[t]
\centering
  \includegraphics[width=\linewidth]{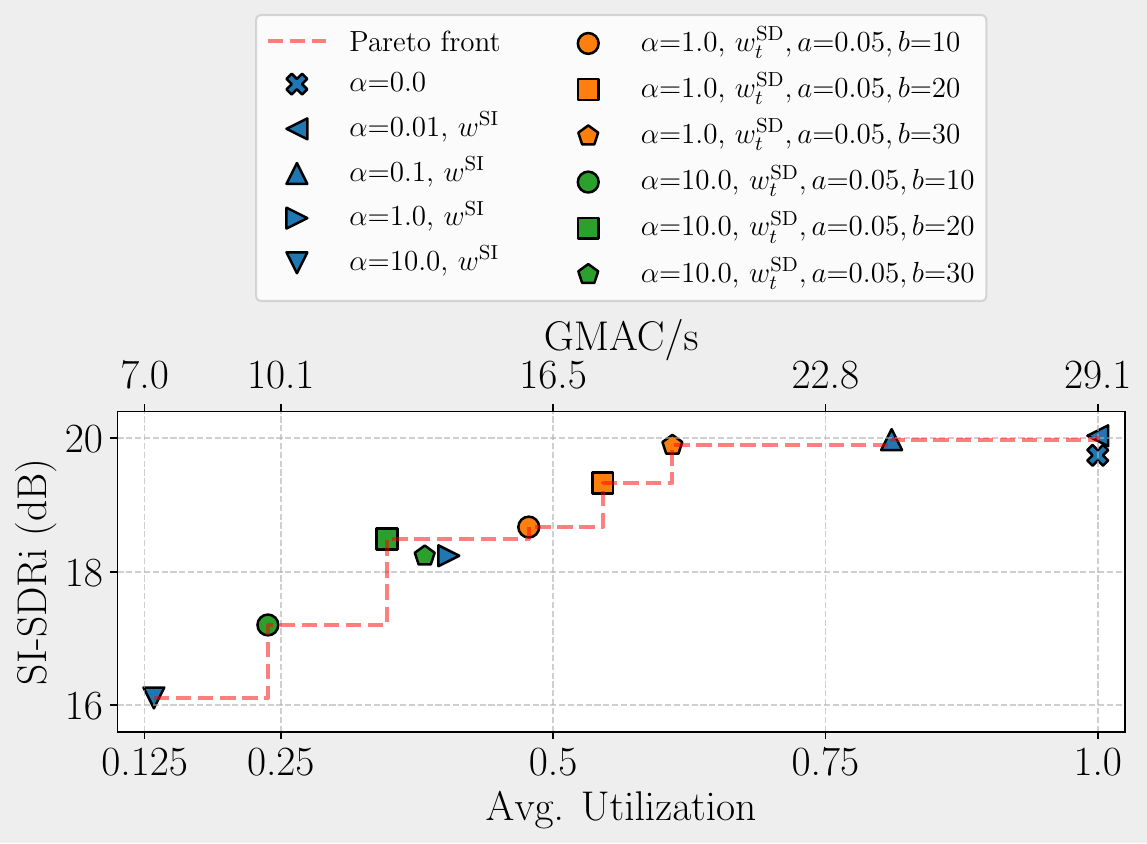} 
  \caption{Trade-off between separation performance and average utilization for different configurations of the complexity loss for the \ac{DSN} ($\mathcal{U}=\{0.125, 1.0\}$). Each point represents the performance of a model averaged over speech mixtures with different overlap ratios.}
  \label{fig:complexity_loss_ablation}
\end{figure}

The training objective for the proposed \ac{DSN} is to simultaneously improve signal reconstruction while operating at reduced computational complexity. The latter objective is enforced by the complexity loss, defined in (\ref{eq:complexity}), to penalize higher utilization. To study its impact, we varied the scaling factor $\alpha$ in (\ref{eq:overall_loss}), which controls the trade-off between reconstruction quality and computational cost. A higher value of $\alpha$ encourages the gating module in the \ac{DSN} to predict lower utilization. Additionally, we examined the impact of the \ac{SD} weighting $w_t$, defined in~(\ref{eq:sd_wf_eq}), using different values of the hyperparameter $b$. Here, we consider the \ac{DSN} with two utilization levels ($\mathcal{U}=\{0.125, 1.0\}$). Figure~\ref{fig:complexity_loss_ablation} summarizes the results in terms of the average \ac{SI-SDR} improvement (\ac{SI-SDR}i) scores and average utilization. The evaluation was performed over speech mixtures with different overlap ratios $\{100\%, 75\%, 50\%, 25\%\}$ 
to account for varying levels of separation difficulty. 
For each model, we report the average performance across the different overlap ratios. The Pareto front is also depicted (red dashed line) to highlight models that achieve the best trade-off between separation performance and computational complexity. 

\begin{figure}[t]
\centering
  \includegraphics[width=\linewidth]{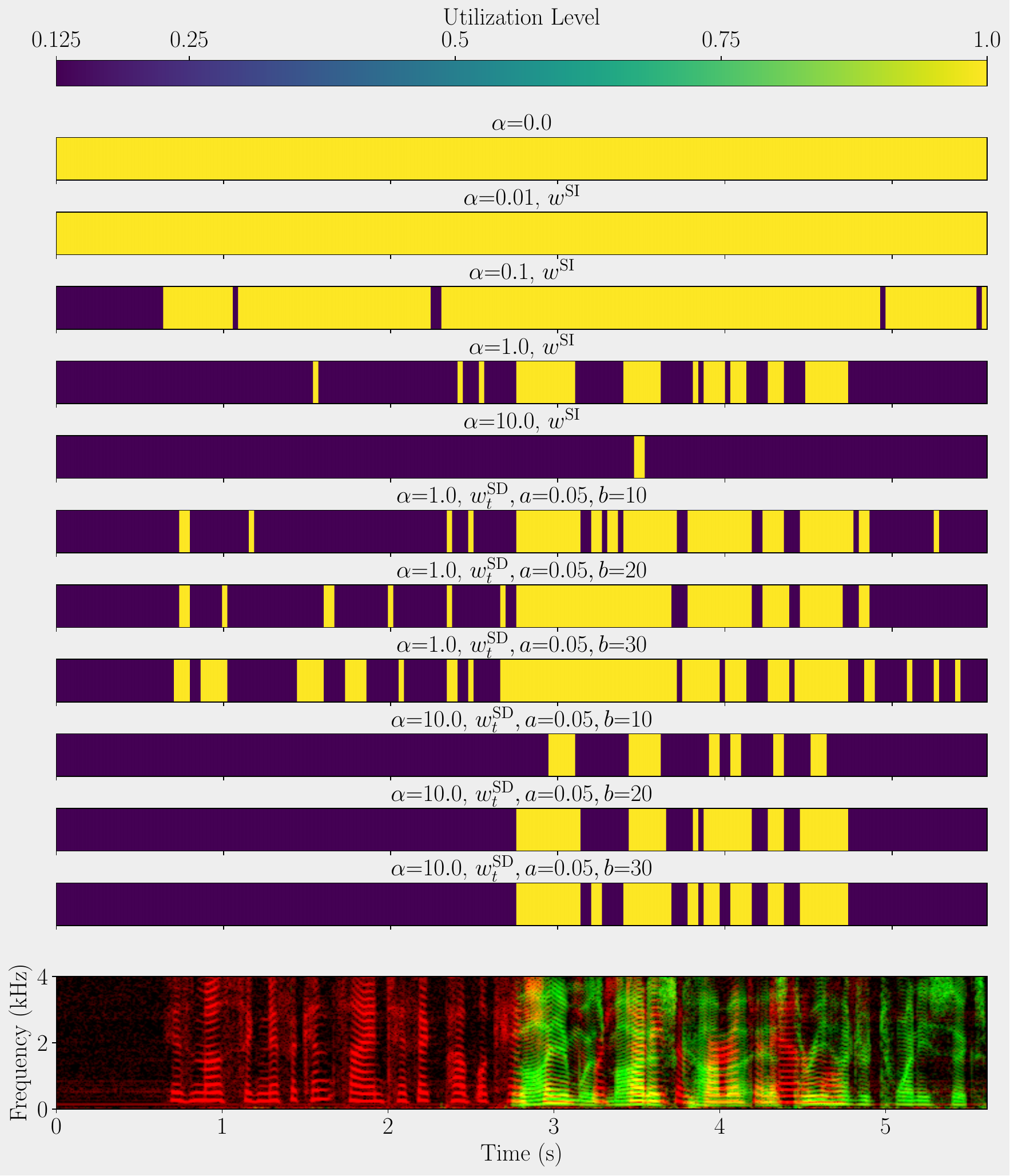} 
  \caption{Predicted utilization over time for different configurations of the complexity loss. The bottom plot depicts the spectrogram of the input two-speaker mixture, with individual speakers highlighted in green and red, and overlapping regions indicated in yellow.}
  \label{fig:visualization_complexity_loss_ablation}
\end{figure}

For models with \ac{SI} weighting (i.e., $w^{\text{SI}}$, shown with blue markers), it can be observed that increasing $\alpha$ reduces the average utilization at the cost of separation performance. When the complexity loss is not applied (i.e., $\alpha=0.0$) or a low value of $\alpha$ is used (e.g., 0.01), the models converge toward static behavior with maximum utilization (i.e., $U_t = 1.0\, \forall t$). This indicates that the gradients from the signal reconstruction loss tend to drive the gating module to predict higher utilization to maximize the separation performance. Conversely, a high value of $\alpha$ (e.g., 10.0) drives the average utilization to its lowest value ($U_t = 0.125\, \forall t$) but noticeably degrades the separation quality. A moderate setting, such as $\alpha = 1.0$, yields a balanced trade-off between computational complexity and separation performance.

Optimizing the \acp{DSN} using the \ac{SD} weighting mechanism (i.e., $w_t^{\text{SD}}$, shown with orange and green markers) enables more flexible control over the performance-complexity trade-off than \ac{SI} weighting. Instead of applying a uniform penalty across all time frames, the \ac{SD} weighting adapts the penalization based on the segmental reconstruction quality, as described in Section~\ref{sec:sd_wf}. A higher value of the threshold parameter $b$ reduces penalization for more challenging segments, which generally results in higher average utilization and separation performance.

Figure~\ref{fig:visualization_complexity_loss_ablation} shows the predicted utilization over time for different models on a test example. For models with \ac{SI} weighting, the utilization is either consistently high or uniformly low across the entire signal, except for $\alpha=0.1$ and $\alpha=1.0$, where dynamic behavior is observed. In contrast, models with \ac{SD} weighting exhibit pronounced temporal adaptation, where higher utilization is used during overlapping speech segments (highlighted in yellow in the spectrogram), and lower utilization during non-overlapping or silent segments. 

In summary, the findings highlight that the complexity loss, particularly when used with \ac{SD} weighting, enables the \ac{DSN} to adapt its computational complexity based on the local characteristics of the input signal.

\subsection{Effect of a Larger Utilization Set}
\begin{figure}[t]
\centering
\includegraphics[width=\linewidth]{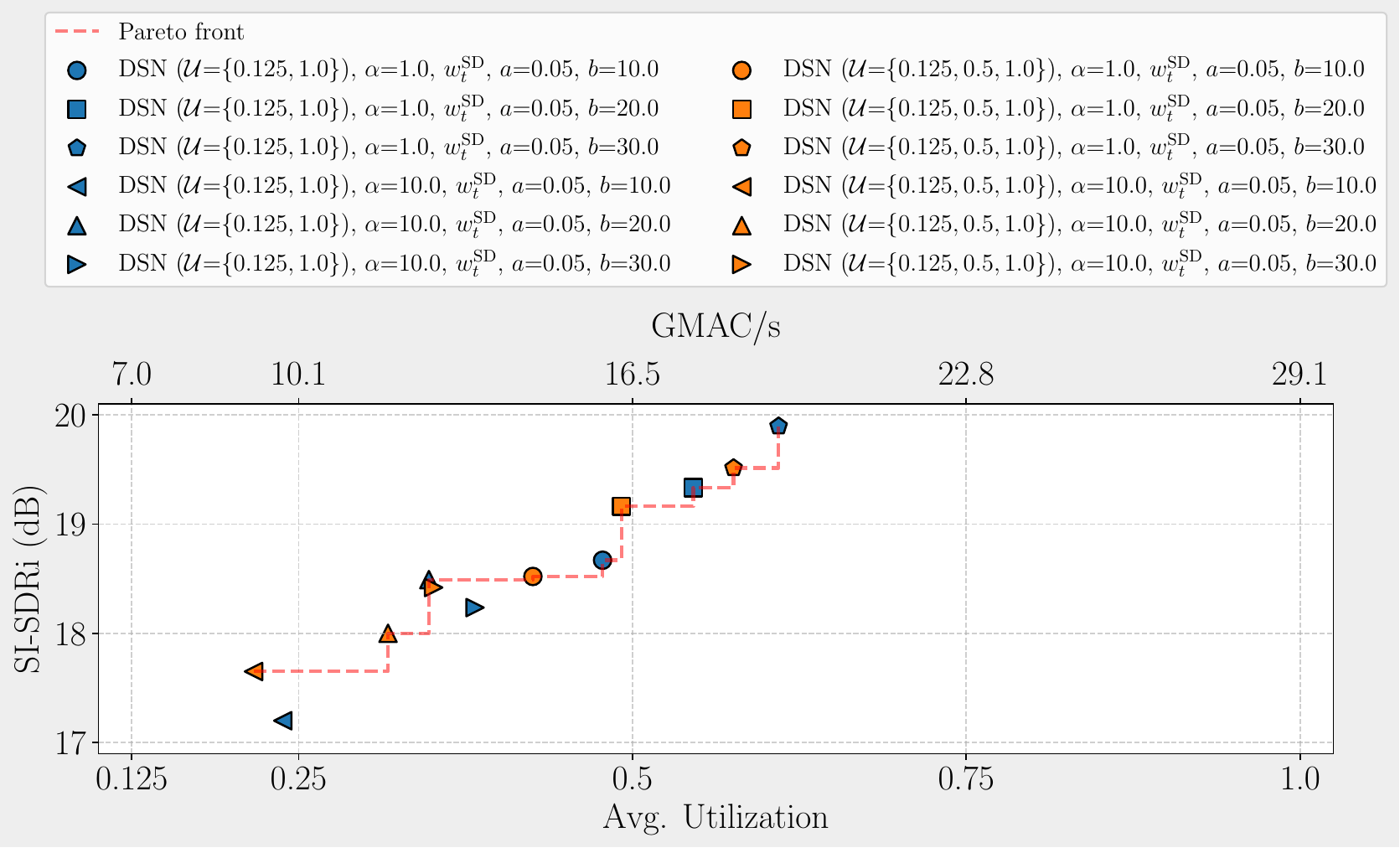} 
    \caption{Trade-off between separation performance and average utilization for different configurations of the complexity loss for two-level and three-level \acp{DSN}. Each point represents the performance of a model averaged over speech mixtures with different overlap ratios.}
\label{fig:complexity_loss_ablation_large_U_set}
\end{figure}

To further demonstrate the flexibility of the proposed \ac{DSN}, we extended the number of utilization levels from two ($\mathcal{U}=\{0.125, 1.0\}$) to three ($\mathcal{U}=\{0.125, 0.5, 1.0\}$). Introducing an intermediate utilization level enables finer control over computational resources, allowing the \ac{DSN} to adapt better to varying input characteristics and thereby improve the computational efficiency. Figure~\ref{fig:complexity_loss_ablation_large_U_set} shows the performance of the two-level and three-level \acp{DSN} for different complexity loss configurations. It can be observed that the three-level \acp{DSN} generally exhibit a better performance-efficiency trade-off than their two-level counterparts, highlighting the effectiveness of a larger utilization set in improving efficiency without compromising separation performance.

\begin{figure}[t]
\centering
  \includegraphics[width=\linewidth]{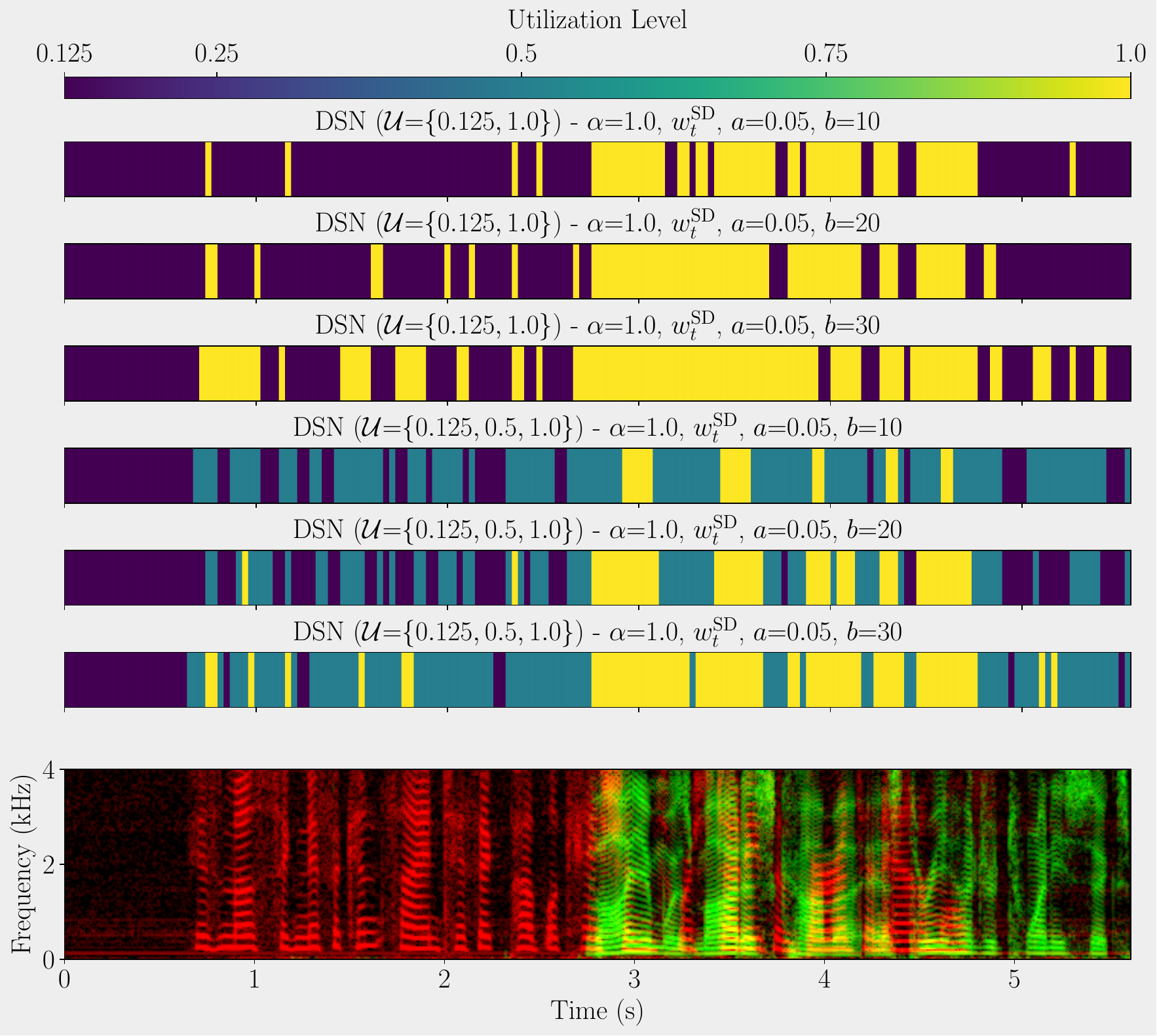} 
  \caption{Predicted utilization over time for two-level and three-level \acp{DSN}, each trained with a different complexity loss configuration. The bottom plot depicts the spectrogram of the input two-speaker mixture, with individual speakers highlighted in green and red, and overlapping regions indicated in yellow.}
\label{fig:visualization_complexity_loss_ablation_large_U_set}
\end{figure}

\begin{figure}[t]
\centering
  \includegraphics[width=\linewidth]{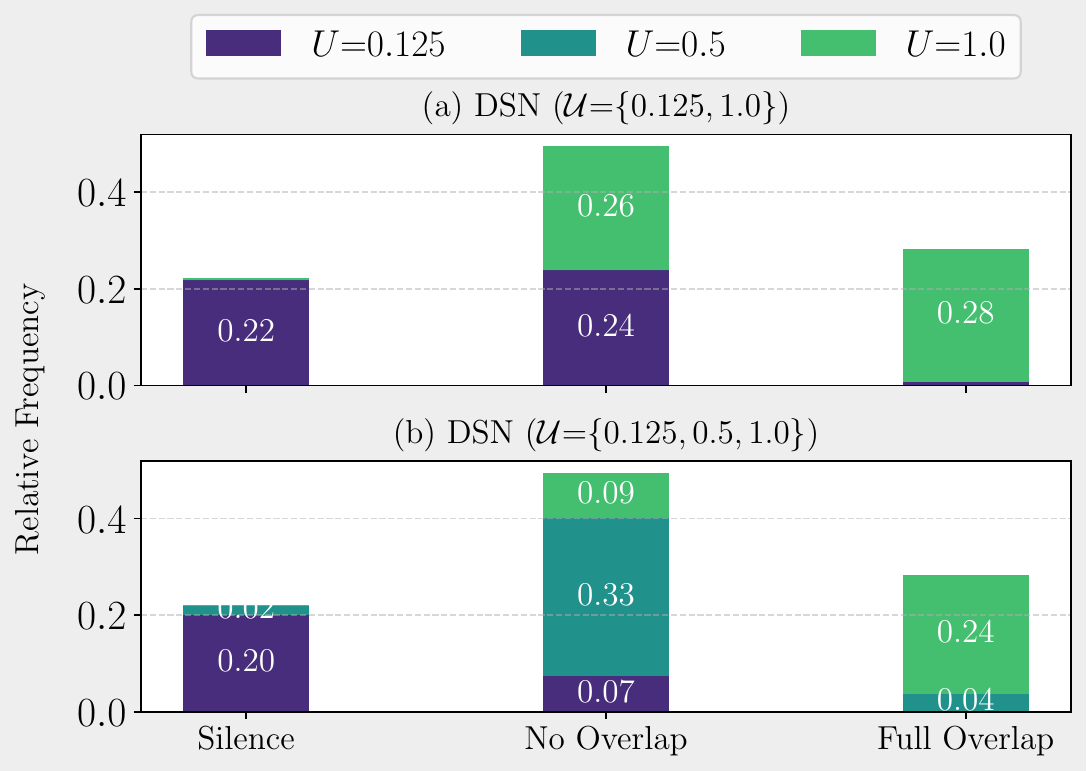} 
  \caption{Distribution of predicted utilization levels with respect to different segment conditions (silence, no overlap, and full overlap) for two-level and three-level \acp{DSN} trained with $\alpha = 1.0$, $w_t^{\text{SD}}$, $a = 0.05$, and $b = 30$.}
  \label{fig:sig_type_utilization}
\end{figure}

Figure~\ref{fig:visualization_complexity_loss_ablation_large_U_set} shows the predicted utilization over time for the two-level and three-level \acp{DSN}, where the latter demonstrate more fine-grained temporal adaptation by frequently selecting the intermediate utilization level during non-overlapping and fully overlapping segments. A quantitative analysis of how the utilization levels are distributed across three input segment types (silence, no overlap, and full overlap) is depicted in Figure~\ref{fig:sig_type_utilization}. Here, we consider two representative models (with $\alpha = 1.0$, $w_t^{\text{SD}}$, $a = 0.05$, and $b = 30$) from the two-level and three-level \acp{DSN} shown in Figure~\ref{fig:complexity_loss_ablation_large_U_set}. For the two-level \ac{DSN}, the lowest and highest utilization levels are primarily used for silent and fully overlapping segments, respectively, while non-overlapping segments show a more balanced distribution between the two levels. The three-level \ac{DSN} further refines this behavior by assigning the intermediate level predominantly to non-overlapping segments, while reserving the lowest and highest levels for silent and fully overlapping segments, respectively. This demonstrates more flexible adaptability to varying signal conditions compared to the two-level \acp{DSN}.  

\subsection{Comparison with Static Networks}
To demonstrate the advantage of dynamic inference, we compared the proposed \ac{DSN} against static networks with fixed computational complexity. The static networks had five distinct model sizes, with computational cost matching that of the \ac{DSN} using fixed utilization levels $U=\{0.125, 0.25, 0.5, 0.75, 1.0\}$. The network structure and training setup for the static networks were identical to those of the \ac{DSN}, except that the number of active heads in the \ac{MHA} and the number of neurons in the intermediate layer of the \ac{FFW} blocks were fixed during both training and inference. We evaluated both the two-level and three-level \acp{DSN}, trained with $\alpha = 1.0$, $w_t^{\text{SD}}$, $a = 0.05$, and $b = 30$.

\begin{figure}[t]
\centering
    \includegraphics[width=\linewidth]{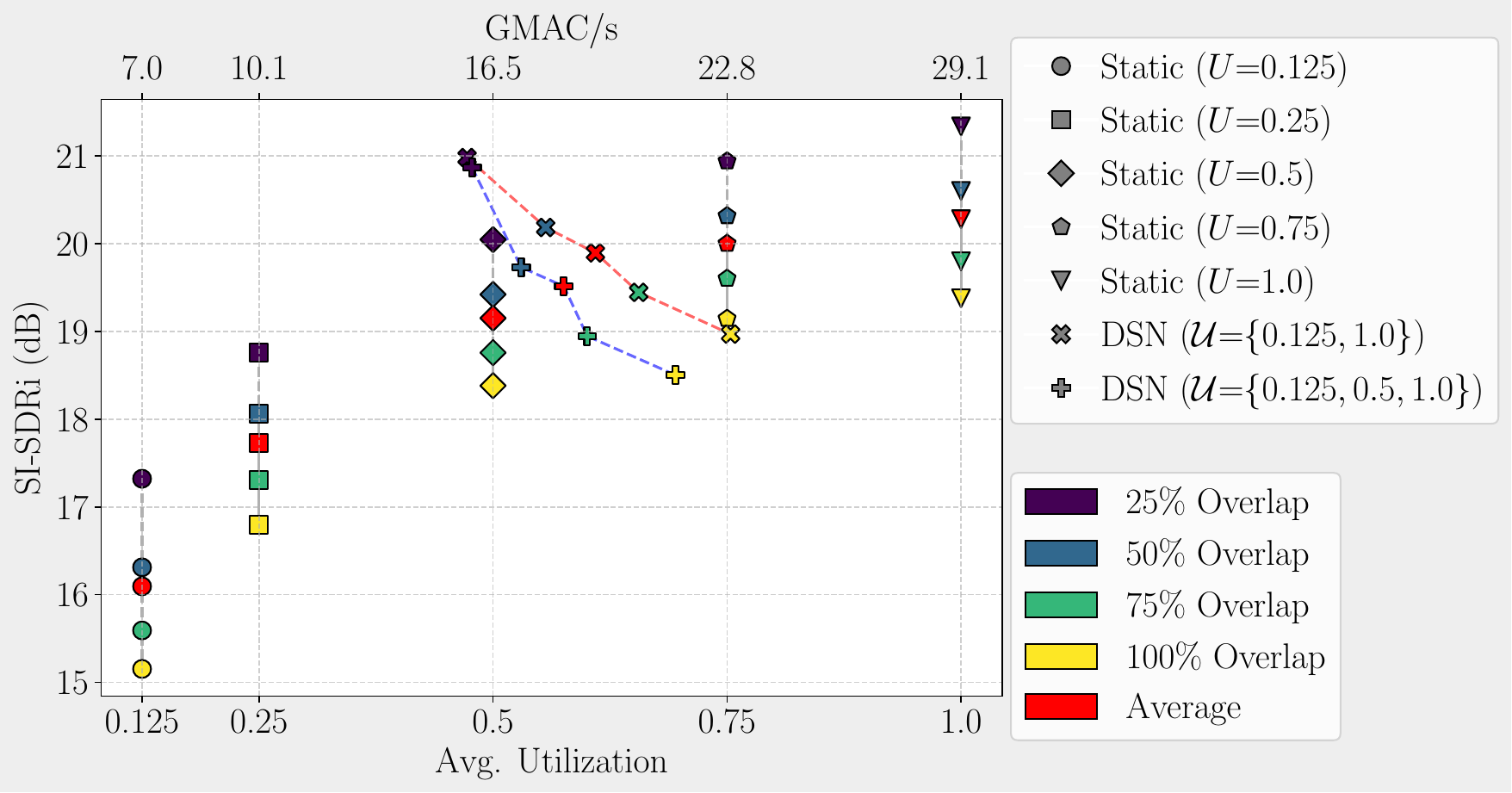} 
    \caption{Performance of the \acp{DSN} compared to static networks with different model sizes. Each marker corresponds to a specific model, while colors represent different overlap ratios. The red markers show the average performance across overlap conditions, while the GMAC/s values on the top axis indicate the corresponding average computational cost. The \acp{DSN} results are shown for models trained with $\alpha = 1.0$, $w_t^{\text{SD}}$, $a = 0.05$, and $b = 30$.}
  \label{fig:static_vs_dynamic}
\end{figure}

Figure~\ref{fig:static_vs_dynamic} presents the separation performance and average utilization across different overlap ratios. For static networks, the performance on fully overlapping mixtures improves monotonically with the model size, showing a gap of approximately 4~dB between the smallest and largest models. Moreover, the performance improves as the overlap ratio decreases, because the separation task generally becomes less challenging in such conditions. In contrast to static networks, the \acp{DSN} dynamically adapt their computational cost based on the characteristics of the input signal, with reduced average utilization observed for mixtures with lower overlap. Notably, the two-level \ac{DSN} achieves average performance comparable to the largest static network while operating at substantially reduced complexity. This efficiency gain is particularly evident at an overlap of $25\%$, where complexity is reduced by over $50\%$ without compromising performance. The three-level \ac{DSN} enables further complexity reduction compared to the two-level variant, but this results in a slight degradation in separation quality.

\begin{figure}[t]
\centering
  \includegraphics[width=\linewidth]{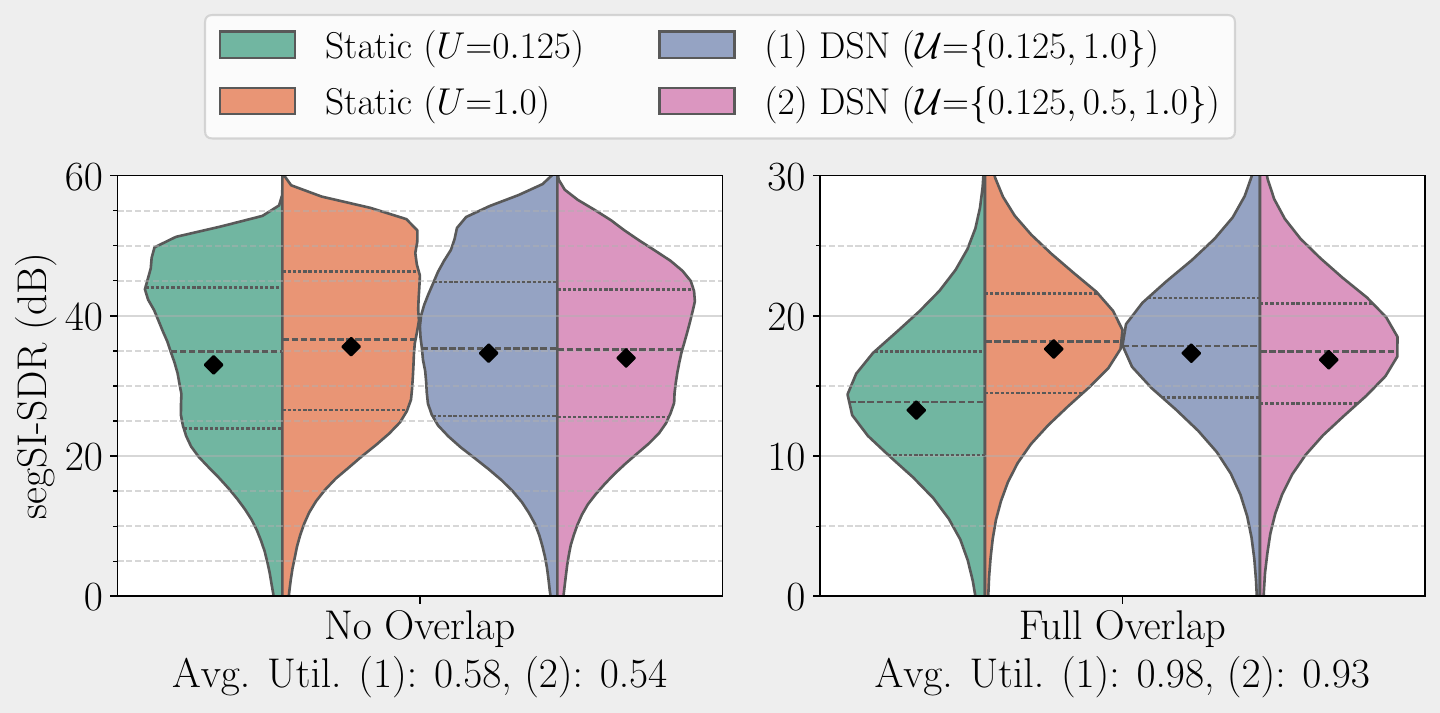} 
  \caption{Distribution of segSI-SDR scores for non-overlapping and fully overlapping segments. The black diamonds indicate the mean segSI-SDR for each model. For the \acp{DSN}, the average utilization is reported.}
  \label{fig:sig_type_segSISDR}
\end{figure}

\begin{figure*}[t]
    \centering
    \subfloat[Non-overlapping segments\label{fig:seg_sisdr_hist2d_partial}]{%
        \includegraphics[width=0.95\linewidth]{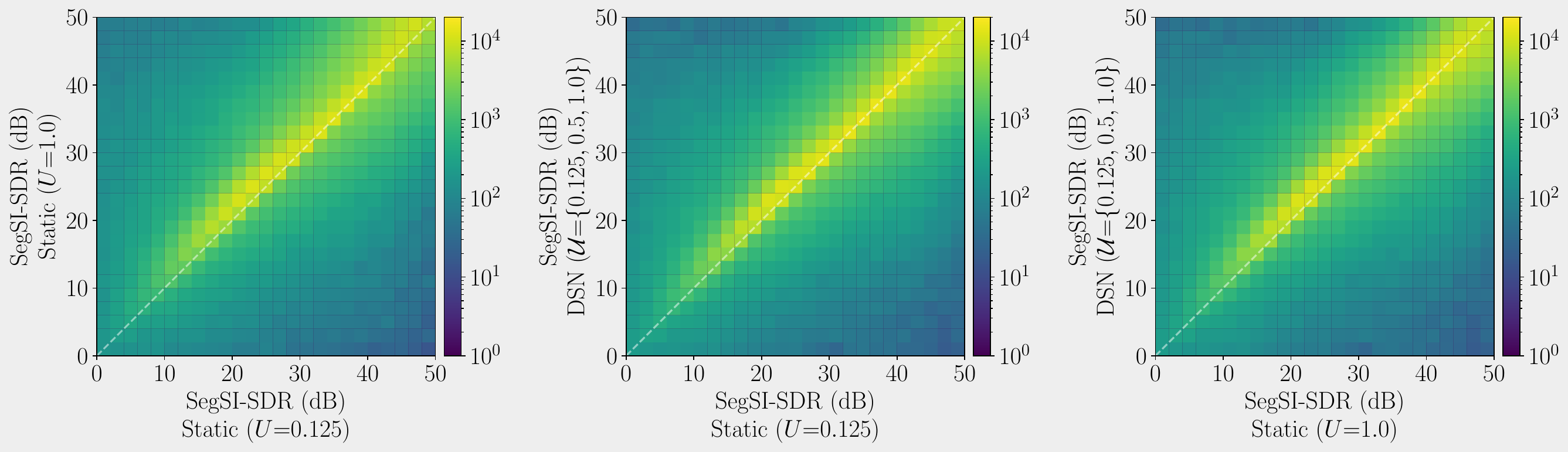}}%
        \\
    \subfloat[Fully overlapping segments\label{fig:seg_sisdr_hist2d_full}]{%
        \includegraphics[width=0.95\linewidth]{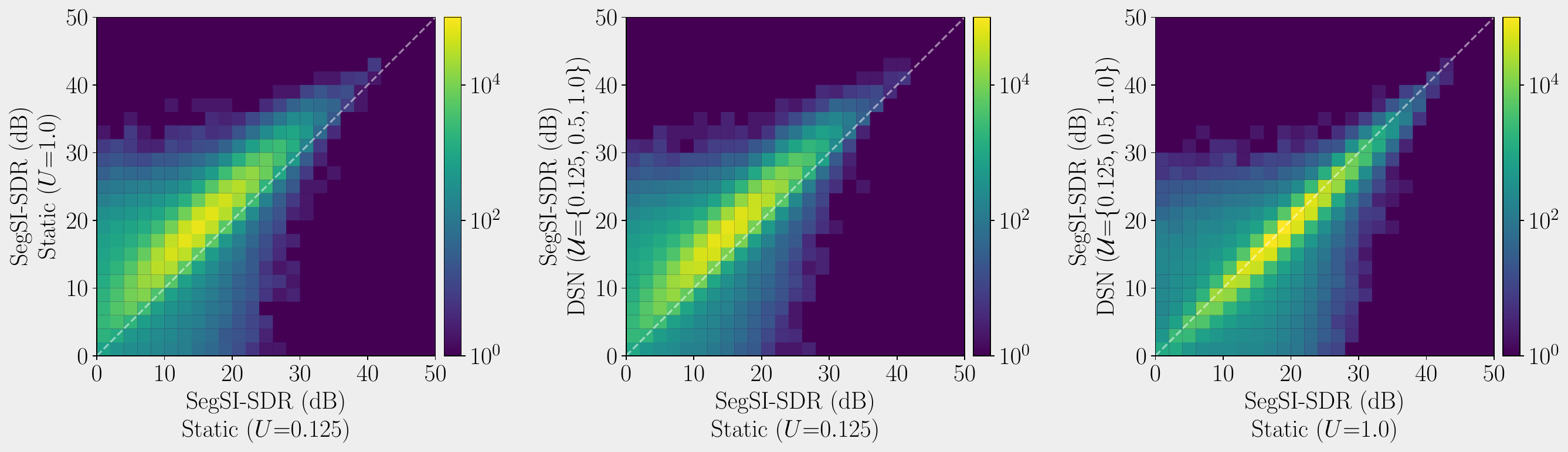}}%
    \caption{Pairwise comparison of segSI-SDR scores between static networks and the \ac{DSN} ($\mathcal{U}= \{0.125, 0.5, 1.0\}$) for (a) non-overlapping segments and (b) fully overlapping segments.}
    \label{fig:seg_sisdr_hist2d}
\end{figure*}

To better understand the performance of the different models under varying signal conditions, we analyze the distribution of segSI-SDR for non-overlapping and fully overlapping segments, as shown in Figure~\ref{fig:sig_type_segSISDR}. For non-overlapping segments, both the smallest and largest static networks achieve high mean and median segSI-SDR scores, indicating that computational savings can be realized without compromising performance for such segments. The two-level and three-level \acp{DSN} maintain high performance in non-overlapping segments, while requiring significantly less computation than the largest static network. For fully overlapping segments, there is a clear performance gap between the smallest and largest static networks. The \acp{DSN} adaptively increase their utilization in such challenging segments, achieving performance comparable to the largest static network. These observations are further supported by the pairwise heatmaps in Figure~\ref{fig:seg_sisdr_hist2d}. The left and middle columns reveal a clear performance gap between the smallest static network and both the largest static network and the \ac{DSN} for fully overlapping segments, while the gap is less pronounced for non-overlapping segments. In the right column, the majority of points cluster around the diagonal, indicating that the \ac{DSN} achieves segSI-SDR scores comparable to those of the largest static network.

\subsection{Robustness to External Gating}
\label{sec:robustness}
In this experiment, we assessed the robustness of the \ac{DSN} when its internal gating mechanism, defined in~(\ref{eq:internal_gating}), is overridden by externally provided gating decisions, a setting referred to as \emph{external gating}. This scenario is particularly relevant for applications where the gating decisions may be determined by external factors, such as the battery level of a wearable device.  We examined both two-level and three-level \acp{DSN}, trained with and without gating dropout (Section~\ref{sec:training_setup}). The results are summarized in Table~\ref{tab:robustness}.

It can be seen that the \acp{DSN} trained without gating dropout exhibit a significant drop in performance when external gating is applied, particularly at low utilization (i.e., $U{=}0.125$). In contrast, incorporating gating dropout during training improves robustness to external gating and reduces the performance gap relative to static networks. These results highlight the effectiveness of using gating dropout in mitigating the performance loss introduced by external gating.

\begin{table}[t]
    \centering
    \setlength{\tabcolsep}{4pt}
    \renewcommand{\arraystretch}{1.7}
    \caption{Performance of the \acp{DSN} under internal and external gating, reported in terms of \ac{SI-SDR}i (dB). For internal gating, the average utilization is shown in parentheses  next to the \ac{SI-SDR}i scores. The \acp{DSN} results are shown for models trained with $\alpha = 1.0$, $w_t^{\text{SD}}$, $a = 0.05$, and $b = 20$.}
    \label{tab:robustness}
    \resizebox{\columnwidth}{!}{%
    \begin{tabular}{@{}lccccc@{}}
    \toprule
    \textbf{Model} & \textbf{Dropout} & \multicolumn{1}{c}{\textbf{Internal Gating}} & \multicolumn{3}{c}{\textbf{External Gating}} \\
                   &                  &           & $U_t{=}0.125$ & $U_t{=}0.5$ & $U_t{=}1.0$ \\ \midrule
    Static & - & -     & 16.1 & 19.2 & 20.3 \\ 
    DSN ($\mathcal{U}{=}\{0.125, 1.0\}$)         & \xmark & 19.7 (0.55) & \phantom{0}7.4 & -    & 16.6 \\
    DSN ($\mathcal{U}{=}\{0.125, 1.0\}$)         & \cmark & 19.3 (0.55) & 15.3 & -    & 19.9 \\
    DSN ($\mathcal{U}{=}\{0.125, 0.5, 1.0\}$)    & \xmark & 19.2 (0.54) & \phantom{0}8.9 & 15.3 & 13.0 \\
    DSN ($\mathcal{U}{=}\{0.125, 0.5, 1.0\}$)    & \cmark & 19.2 (0.49) & 15.0 & 18.6 & 19.5 \\ 
    \bottomrule  
    \end{tabular}%
    }
\end{table} 
\begin{figure*}[t]
\centering
    \includegraphics[width=\linewidth]{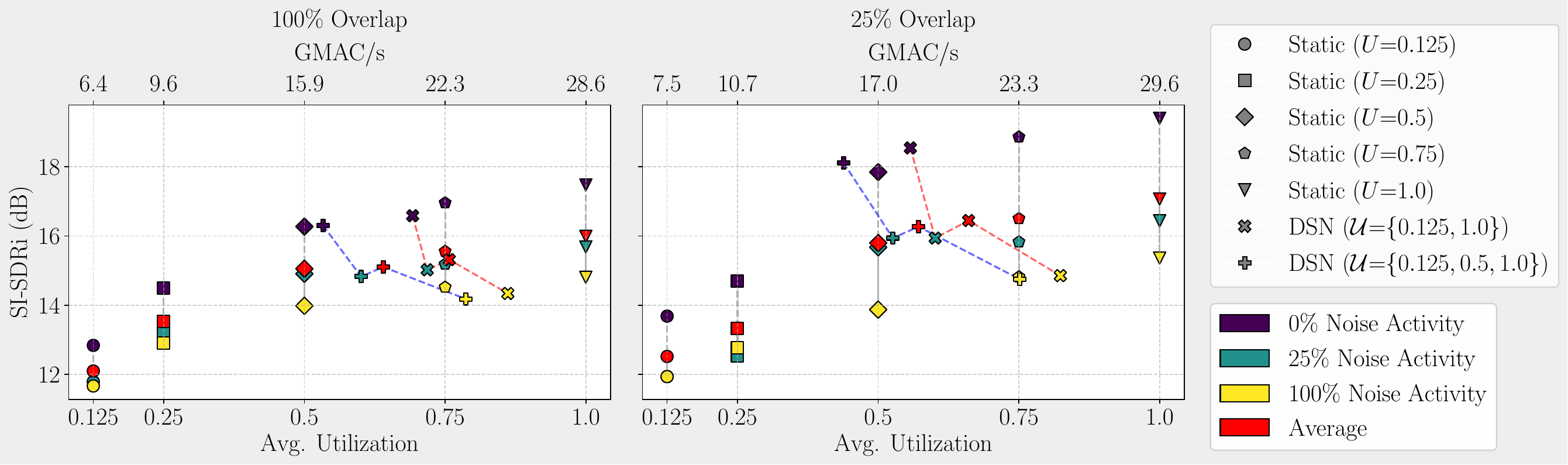} 
    \caption{Performance of the \acp{DSN} compared to static networks on speech mixtures from WHAM~\cite{wichern2019wham} with different noise activities for fully and partially overlapping mixtures in (a) and (b), respectively. Each marker corresponds to a specific model, while colors represent different noise activity levels. Each red marker indicates the average performance across the different noise activities. The \acp{DSN} results are shown for models trained with $\alpha = 1.0$, $w_t^{\text{SD}}$, $a = 0.05$, and $b = 10$.}
  \label{fig:static_vs_dynamic_whamr}
\end{figure*}

\subsection{Performance on WHAM! Dataset}
To evaluate the proposed \ac{DSN} on more realistic input conditions, we performed experiments on the WHAM! dataset~\cite{wichern2019wham}, which adds background noise to the mixtures in the WSJ0-2mix dataset, resulting in a more challenging separation task. Static networks with different model sizes and \acp{DSN} with two and three utilization levels were trained and evaluated on noisy mixtures with varying noise activities and overlap ratios. The results are shown in Figure~\ref{fig:static_vs_dynamic_whamr}. 

For fully overlapping mixtures (Figure~\ref{fig:static_vs_dynamic_whamr}a), static networks with higher utilization levels yield better separation performance, and the models generally perform better as the noise activity decreases. The \acp{DSN} exhibit dynamic behavior by adjusting their utilization in response to the noise activity level. However, they do not achieve a more favorable performance-efficiency trade-off compared to static networks with $U{=}0.5$ or $U{=}0.75$. In contrast, for partially overlapping mixtures (Figure~\ref{fig:static_vs_dynamic_whamr}b), the \acp{DSN} demonstrate a clearer advantage over the static networks, particularly under lower noise activity levels.

\begin{figure}[t]
    \centering
    \subfloat[Noisy mixture\label{fig:visualization_utilization_whamr_AN}]{%
        \includegraphics[width=\linewidth]{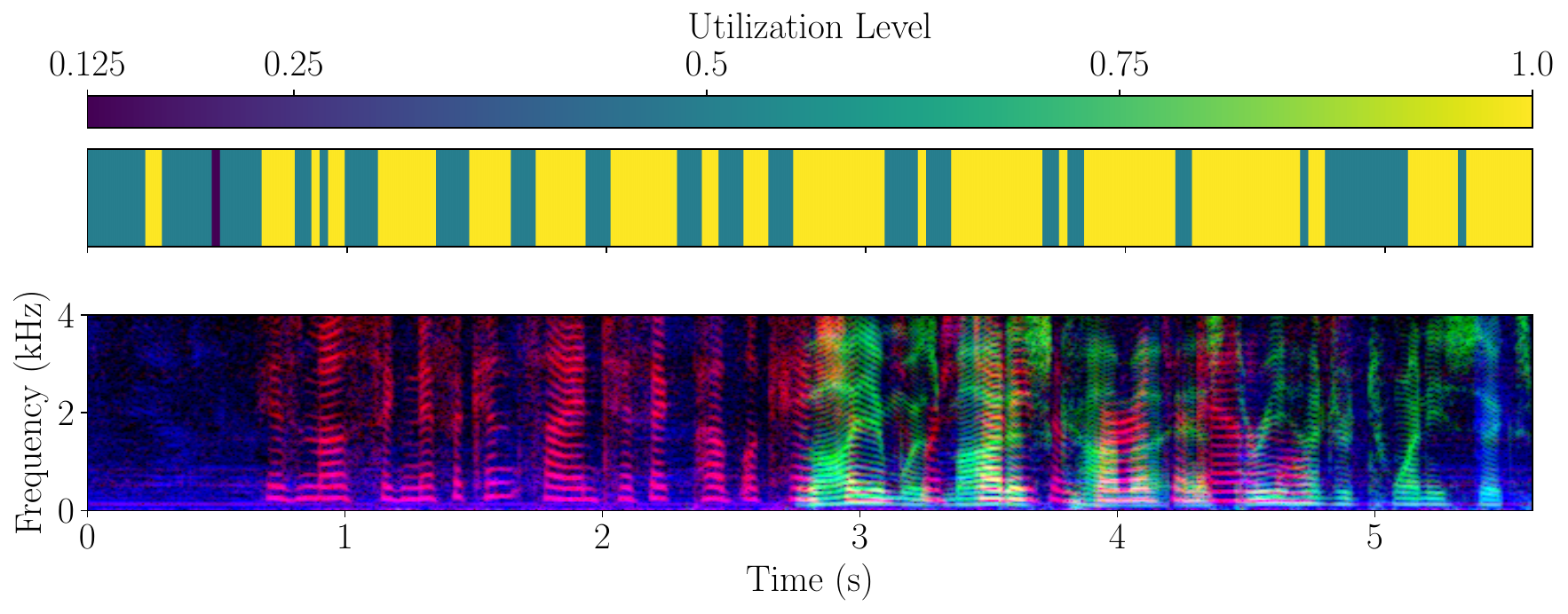}}%
    \\
    \subfloat[Clean mixture\label{fig:visualization_utilization_whamr_A}]{%
        \includegraphics[width=\linewidth]{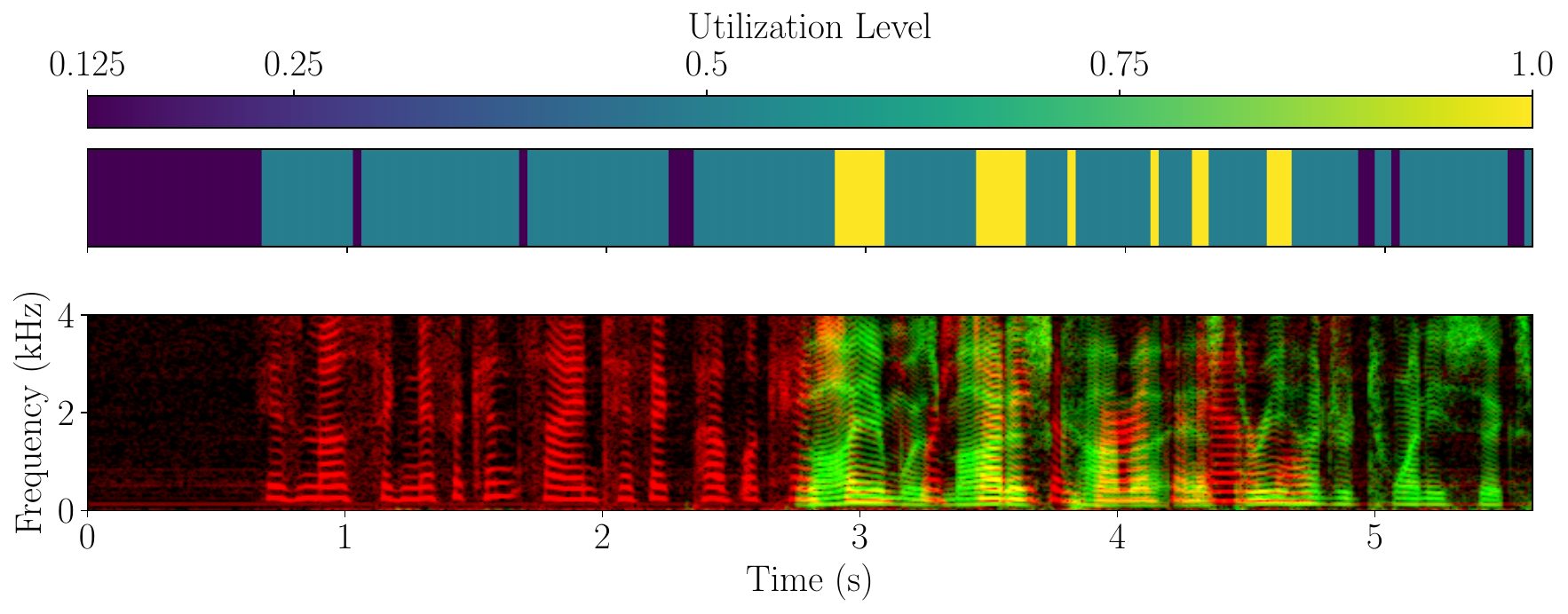}}%
    \caption{Predicted utilization over time for two WHAM! test examples: (a) noisy mixture and (b) clean mixture. The blue regions in the spectrogram in (a) correspond to the background noise.}
    \label{fig:visualization_utilization_whamr}
\end{figure}

\begin{figure}[t]
\centering
  \includegraphics[width=\linewidth]{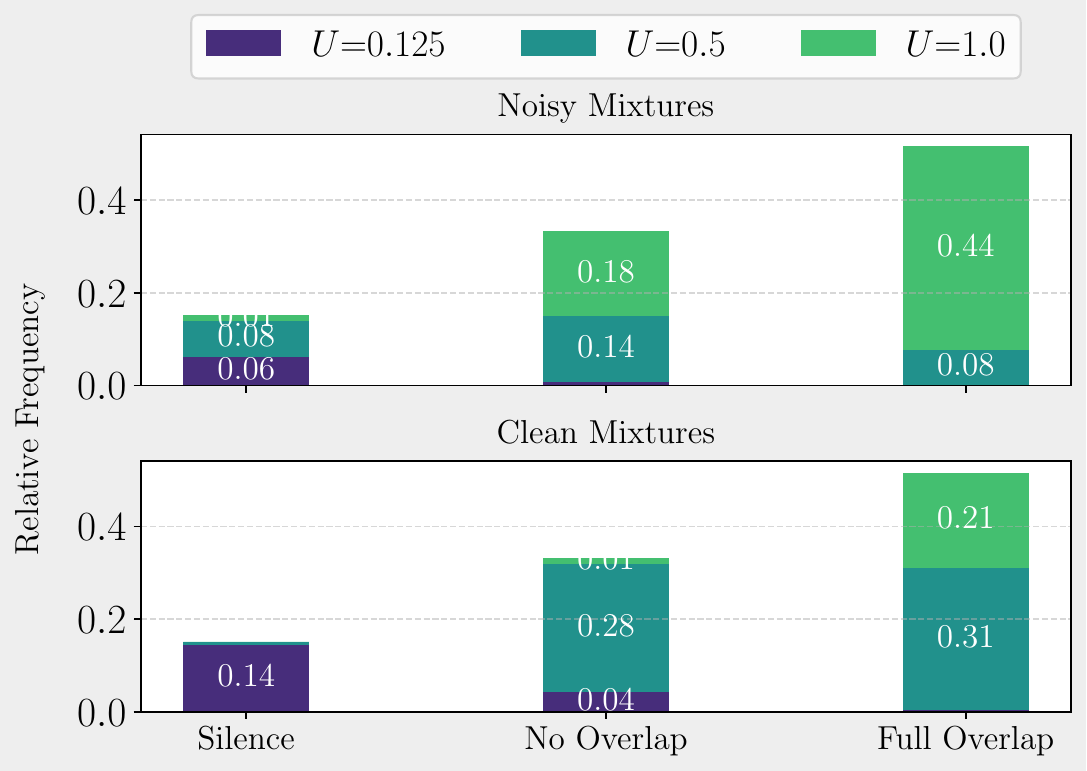} 
  \caption{Distribution of predicted utilization levels by the \ac{DSN} ($\mathcal{U}=\{0.125, 0.5, 1.0\}$) with respect to different segments conditions (silence, no overlap, and full overlap) for noisy mixtures (top) and clean mixtures (bottom).}
  \label{fig:sig_type_utilization_whamr}
\end{figure}

Figure~\ref{fig:visualization_utilization_whamr} shows the predicted utilization levels by the \ac{DSN} ($\mathcal{U}= \{0.125, 0.5, 1.0\}$) for a noisy and a clean mixture. The presence of background noise in the mixture (Figure~\ref{fig:visualization_utilization_whamr_AN}) leads to higher utilization compared to the clean case (Figure~\ref{fig:visualization_utilization_whamr_A}), demonstrating the \ac{DSN}'s effective adaptation not only to the speech activity but also to the noise condition.
Figure~\ref{fig:sig_type_utilization_whamr} further quantifies this behavior by analyzing the distribution of predicted utilization levels across the different segment types. In both noisy and clean conditions, the lowest utilization level (i.e., $U{=}0.125$) is mostly assigned to silent segments, while the highest utilization level (i.e., $U{=}1.0$) is increasingly favored in fully overlapping segments. Notably, the shift toward higher utilization is more pronounced in the noisy case, with $U{=}1.0$  being more frequently selected even in non-overlapping segments, highlighting the increased computational cost imposed by the background noise.

\begin{figure}[t]
\centering
  \includegraphics[width=\linewidth]{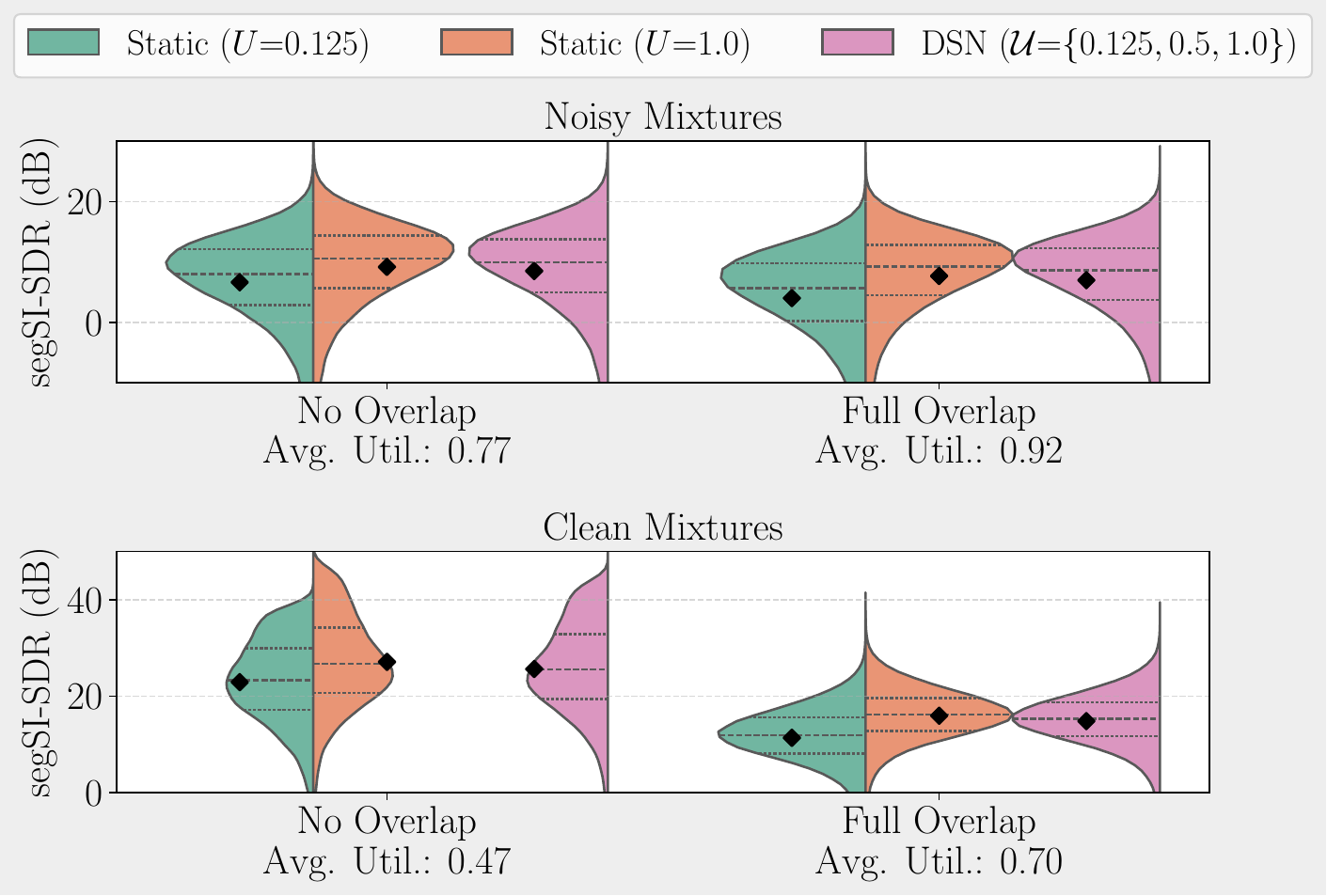} 
  \caption{Distribution of segSI-SDR scores for non-overlapping and fully overlapping segments. Performance on noisy mixtures (top) and clean mixtures (bottom). The mean segSI-SDR  scores are indicated by the black diamonds. The average utilization of the \ac{DSN} for each condition is reported.}
  \label{fig:sig_type_segSISDR_whamr}
\end{figure}

Figure~\ref{fig:sig_type_segSISDR_whamr} presents the distribution of segSI-SDR scores for non-overlapping and fully overlapping segments in noisy and clean mixtures. The \ac{DSN} consistently outperforms the smallest static network ($U{=}0.125$) across all conditions and approaches the performance of the largest static network ($U{=}1.0$), despite operating at a reduced average utilization. These results confirm the effectiveness of the \ac{DSN} in reducing the computational cost without significantly compromising the separation quality, even under challenging noisy conditions.

\section{Discussion}
\label{sec:discussion}
\acresetall
In this paper, we presented a \ac{DSN} for \acl{SS} that enables frame-level adaptation of computational complexity based on local characteristics of the input signal. The \ac{DSN} was optimized to simultaneously improve signal reconstruction while promoting computational efficiency. To achieve a better performance-complexity trade-off, we proposed a \acl{SD} weighting scheme in the complexity loss, where the degree of penalization depends on the network's performance, as measured by the segmental reconstruction quality. This encourages the allocation of more computations to challenging segments (e.g., fully overlapping speech) and fewer to simpler ones (e.g., silence or non-overlapping speech). Our experiments on the WSJ0-2mix and WHAM! datasets demonstrate that the \ac{DSN} achieves a favorable trade-off between separation performance and computational efficiency compared to individually trained static networks of different model sizes. 
Especially under low-overlap conditions, the \ac{DSN} achieves similar separation quality to static networks while requiring fewer computations, making it well-suited for real-world scenarios such as natural conversations, where speaker overlap typically remains below $20\%$~\cite{cetin2006analysis}.

Overall, our findings highlight the potential of \acp{DSN} as a promising paradigm for efficient \acl{SS} and speech processing systems more broadly. While our work focuses on dynamic width scaling via network slimming, it could be combined with other dynamic inference approaches, such as layer skipping, to provide additional flexibility in controlling the computational complexity. 
However, a key limitation of dynamic inference in general is the lack of hardware support, as existing platforms are mostly optimized for static computation graphs. Translating the model's dynamic behavior into real-world speed-ups thus remains an open challenge and requires further development.

\section*{Acknowledgments}
The authors thank the Erlangen Regional Computing Center (RRZE) for providing compute resources and technical support.

\bibliographystyle{IEEEtran}
\bibliography{sapref}

\begin{thebibliography}{10}
\providecommand{\url}[1]{#1}
\csname url@samestyle\endcsname
\providecommand{\newblock}{\relax}
\providecommand{\bibinfo}[2]{#2}
\providecommand{\BIBentrySTDinterwordspacing}{\spaceskip=0pt\relax}
\providecommand{\BIBentryALTinterwordstretchfactor}{4}
\providecommand{\BIBentryALTinterwordspacing}{\spaceskip=\fontdimen2\font plus
\BIBentryALTinterwordstretchfactor\fontdimen3\font minus \fontdimen4\font\relax}
\providecommand{\BIBforeignlanguage}[2]{{%
\expandafter\ifx\csname l@#1\endcsname\relax
\typeout{** WARNING: IEEEtran.bst: No hyphenation pattern has been}%
\typeout{** loaded for the language `#1'. Using the pattern for}%
\typeout{** the default language instead.}%
\else
\language=\csname l@#1\endcsname
\fi
#2}}
\providecommand{\BIBdecl}{\relax}
\BIBdecl

\bibitem{xu2014regression}
Y.~Xu, J.~Du, L.-R. Dai, and C.-H. Lee, ``A regression approach to speech enhancement based on deep neural networks,'' \emph{{IEEE/ACM} Trans. Audio, Speech, Lang. Process.}, vol.~23, no.~1, pp. 7--19, Jan. 2015.

\bibitem{zheng2023sixty}
C.~Zheng, H.~Zhang, W.~Liu, X.~Luo, A.~Li, X.~Li, and B.~C.~J. Moore, ``Sixty years of frequency-domain monaural speech enhancement: From traditional to deep learning methods,'' \emph{Trends in Hearing}, vol.~27, pp. 1--52, Nov. 2023.

\bibitem{nassif2019speech}
A.~B. Nassif, I.~Shahin, I.~Attili, M.~Azzeh, and K.~Shaalan, ``Speech recognition using deep neural networks: A systematic review,'' \emph{IEEE Access}, vol.~7, pp. 19\,143--19\,165, Feb. 2019.

\bibitem{prabhavalkar2023end}
R.~Prabhavalkar, T.~Hori, T.~N. Sainath, R.~Schlüter, and S.~Watanabe, ``End-to-end speech recognition: A survey,'' \emph{{IEEE/ACM} Trans. Audio, Speech, Lang. Process.}, vol.~32, pp. 325--351, Oct. 2024.

\bibitem{lopez2021deep}
I.~López-Espejo, Z.-H. Tan, J.~H.~L. Hansen, and J.~Jensen, ``Deep spoken keyword spotting: An overview,'' \emph{IEEE Access}, vol.~10, pp. 4169--4199, Dec. 2022.

\bibitem{wang2018supervised}
D.~Wang and J.~Chen, ``Supervised speech separation based on deep learning: {A}n overview,'' \emph{{IEEE/ACM} Trans. Audio, Speech, Lang. Process.}, vol.~26, no.~10, pp. 1702--1726, Oct. 2018.

\bibitem{Luo2019a}
Y.~Luo, Z.~Chen, and T.~Yoshioka, ``Dual-path {RNN}: {E}fficient long sequence modeling for time-domain single-channel speech separation,'' in \emph{Proc. {IEEE} Intl. Conf. on Acoustics, Speech and Signal Processing (ICASSP)}, May 2020, pp. 46--50.

\bibitem{subakan2021attention}
C.~Subakan, M.~Ravanelli, S.~Cornell, M.~Bronzi, and J.~Zhong, ``Attention is all you need in speech separation,'' in \emph{Proc. {IEEE} Intl. Conf. on Acoustics, Speech and Signal Processing (ICASSP)}, Jun. 2021, pp. 21--25.

\bibitem{subakan2023exploring}
C.~Subakan, M.~Ravanelli, S.~Cornell, F.~Grondin, and M.~Bronzi, ``Exploring self-attention mechanisms for speech separation,'' \emph{{IEEE/ACM} Trans. Audio, Speech, Lang. Process.}, vol.~31, pp. 2169--2180, Jun. 2023.

\bibitem{wang2023tf}
Z.-Q. Wang, S.~Cornell, S.~Choi, Y.~Lee, B.-Y. Kim, and S.~Watanabe, ``{TF-GridNet}: {I}ntegrating full- and sub-band modeling for speech separation,'' \emph{{IEEE/ACM} Trans. Audio, Speech, Lang. Process.}, vol.~31, pp. 3221--3236, Aug. 2023.

\bibitem{chetupalli2023speaker}
S.~R. Chetupalli and E.~A.~P. Habets, ``Speaker counting and separation from single-channel noisy mixtures,'' \emph{{IEEE/ACM} Trans. Audio, Speech, Lang. Process.}, vol.~31, pp. 1681--1692, Apr. 2023.

\bibitem{han2021dynamic}
Y.~Han, G.~Huang, S.~Song, L.~Yang, H.~Wang, and Y.~Wang, ``Dynamic neural networks: A survey,'' \emph{{IEEE} Trans. Pattern Anal. Mach. Intell.}, vol.~44, no.~11, pp. 7436--7456, Nov. 2022.

\bibitem{jacobs1991adaptive}
R.~A. Jacobs, M.~I. Jordan, S.~J. Nowlan, and G.~E. Hinton, ``Adaptive mixtures of local experts,'' \emph{Neural Computation}, vol.~3, no.~1, pp. 79--87, Mar. 1991.

\bibitem{fedus2022switch}
W.~Fedus, B.~Zoph, and N.~Shazeer, ``Switch transformers: Scaling to trillion parameter models with simple and efficient sparsity,'' \emph{Journal of Machine Learning Research}, vol.~23, no.~1, pp. 5232--5270, Apr. 2022.

\bibitem{you2021speechmoe}
Z.~You, S.~Feng, D.~Su, and D.~Yu, ``{SpeechMoE}: {S}caling to large acoustic models with dynamic routing mixture of experts,'' in \emph{Proc. Interspeech Conf.}, Aug. 2021, pp. 2077--2081.

\bibitem{bolukbasi17a}
T.~Bolukbasi, J.~Wang, O.~Dekel, and V.~Saligrama, ``Adaptive neural networks for efficient inference,'' in \emph{Proc. Intl. Conf. Machine Learning ({ICML})}, Aug. 2017, pp. 527--536.

\bibitem{Wang_2018_ECCV}
X.~Wang, F.~Yu, Z.-Y. Dou, T.~Darrell, and J.~E. Gonzalez, ``{SkipNet}: {L}earning dynamic routing in convolutional networks,'' in \emph{Proc. European Conf. on Computer Vision (ECCV)}, Sep. 2018, pp. 409--424.

\bibitem{campos2017skip}
V.~Campos, B.~Jou, X.~Gir{\'{o}}{-}i{-}Nieto, J.~Torres, and S.~Chang, ``Skip {RNN: L}earning to skip state updates in recurrent neural networks,'' in \emph{Proc. {IEEE} Intl. Conf. on Learn. Repr. (ICLR)}, May 2018.

\bibitem{li2021dynamic}
C.~Li, G.~Wang, B.~Wang, X.~Liang, Z.~Li, and X.~Chang, ``Dynamic slimmable network,'' in \emph{Proc. IEEE/CVF Conf. on Computer Vision and Pattern Recognition (CVPR)}, Jun. 2021, pp. 8607--8617.

\bibitem{chen2019you}
Z.~Chen, Y.~Li, S.~Bengio, and S.~Si, ``You look twice: {GaterNet} for dynamic filter selection in {CNNs},'' in \emph{Proc. IEEE/CVF Conf. on Computer Vision and Pattern Recognition (CVPR)}, Jun. 2019, pp. 9172--9180.

\bibitem{hua2019channel}
W.~Hua, Y.~Zhou, C.~M. De~Sa, Z.~Zhang, and G.~E. Suh, ``Channel gating neural networks,'' in \emph{Proc. Neural Information Processing Conf (NeurIPS)}, vol.~32, 2019.

\bibitem{hershey2016deep}
J.~R. Hershey, Z.~Chen, J.~{Le Roux}, and S.~Watanabe, ``Deep clustering: Discriminative embeddings for segmentation and separation,'' in \emph{Proc. {IEEE} Intl. Conf. on Acoustics, Speech and Signal Processing (ICASSP)}, Mar. 2016, pp. 31--35.

\bibitem{wichern2019wham}
G.~Wichern, J.~Antognini, M.~Flynn, L.~R. Zhu, E.~McQuinn, D.~Crow, E.~Manilow, and J.~Le~Roux, ``{WHAM!: E}xtending speech separation to noisy environments,'' in \emph{Proc. Interspeech Conf.}, Sep. 2019.

\bibitem{chen2021don}
S.~Chen, Y.~Wu, Z.~Chen, T.~Yoshioka, S.~Liu, J.~Li, and X.~Yu, ``Don’t shoot butterfly with rifles: Multi-channel continuous speech separation with early exit transformer,'' in \emph{Proc. {IEEE} Intl. Conf. on Acoustics, Speech and Signal Processing (ICASSP)}, Jun. 2021, pp. 6139--6143.

\bibitem{bralios2022latent}
D.~Bralios, E.~Tzinis, G.~Wichern, P.~Smaragdis, and J.~L. Roux, ``Latent iterative refinement for modular source separation,'' in \emph{Proc. {IEEE} Intl. Conf. on Acoustics, Speech and Signal Processing (ICASSP)}, Jun. 2023, pp. 1--5.

\bibitem{le2022inference}
X.~Le, T.~Lei, K.~Chen, and J.~Lu, ``Inference skipping for more efficient real-time speech enhancement with parallel {RNNs},'' \emph{{IEEE/ACM} Trans. Audio, Speech, Lang. Process.}, vol.~30, pp. 2411--2421, Jul. 2022.

\bibitem{miccini2024scalable}
R.~Miccini, C.~Laroche, T.~Piechowiak, and L.~Pezzarossa, ``Scalable speech enhancement with dynamic channel pruning,'' in \emph{Proc. {IEEE} Intl. Conf. on Acoustics, Speech and Signal Processing (ICASSP)}, Apr. 2025, pp. 1--5.

\bibitem{elminshawi2023slimtasnet}
M.~Elminshawi, S.~R. Chetupalli, and E.~A.~P. Habets, ``Slim-{T}as{N}et: {A} slimmable neural network for speech separation,'' in \emph{Proc. {IEEE} Workshop on Applications of Signal Processing to Audio and Acoustics ({WASPAA})}, Oct. 2023, pp. 1--5.

\bibitem{elminshawi2024dynamic}
------, ``Dynamic slimmable network for speech separation,'' \emph{{IEEE} Signal Process. Lett.}, vol.~31, pp. 2205--2209, Aug. 2024.

\bibitem{jang2017categorical}
E.~Jang, S.~Gu, and B.~Poole, ``Categorical reparameterization with gumbel-softmax,'' in \emph{Proc. {IEEE} Intl. Conf. on Learn. Repr. (ICLR)}, Apr. 2017.

\bibitem{bengio2013estimating}
Y.~Bengio, N.~L{\'e}onard, and A.~Courville, ``Estimating or propagating gradients through stochastic neurons for conditional computation,'' Aug. 2013, arXiv:1308.3432.

\bibitem{vaswani2017attention}
A.~Vaswani, N.~Shazeer, N.~Parmar, J.~Uszkoreit, L.~Jones, A.~N. Gomez, L.~u. Kaiser, and I.~Polosukhin, ``Attention is all you need,'' in \emph{Proc. Neural Information Processing Conf (NeurIPS)}, vol.~30, 2017.

\bibitem{jin2024moh}
P.~Jin, B.~Zhu, L.~Yuan, and S.~Yan, ``{MoH: M}ulti-head attention as mixture-of-head attention,'' Oct. 2024, arXiv:2410.11842.

\bibitem{leroux2019}
J.~{Le Roux}, S.~Wisdom, H.~Erdogan, and J.~R. Hershey, ``{SDR}--half-baked or well done?'' in \emph{Proc. {IEEE} Intl. Conf. on Acoustics, Speech and Signal Processing (ICASSP)}, May 2019, pp. 626--630.

\bibitem{kolbaek2017multitalker}
M.~Kolbæk, D.~Yu, Z.-H. Tan, and J.~Jensen, ``Multitalker speech separation with utterance-level permutation invariant training of deep recurrent neural networks,'' \emph{{IEEE/ACM} Trans. Audio, Speech, Lang. Process.}, vol.~25, no.~10, pp. 1901--1913, Oct. 2017.

\bibitem{SB2021}
M.~Ravanelli, T.~Parcollet, P.~Plantinga, A.~Rouhe, S.~Cornell, L.~Lugosch, C.~Subakan, N.~Dawalatabad, A.~Heba, J.~Zhong, J.-C. Chou, S.-L. Yeh, S.-W. Fu, C.-F. Liao, E.~Rastorgueva, F.~Grondin, W.~Aris, H.~Na, Y.~Gao, R.~D. Mori, and Y.~Bengio, ``{SpeechBrain}: A general-purpose speech toolkit,'' Jun. 2021, arXiv:2106.04624.

\bibitem{Kingma2015}
D.~P. Kingma and J.~Ba, ``Adam: {A} method for stochastic optimization,'' in \emph{Proc. {IEEE} Intl. Conf. on Learn. Repr. (ICLR)}, May 2015, pp. 1--15.

\bibitem{cetin2006analysis}
Özgür Çetin and E.~Shriberg, ``Analysis of overlaps in meetings by dialog factors, hot spots, speakers, and collection site: {I}nsights for automatic speech recognition,'' in \emph{Proc. Interspeech Conf.}, Sep. 2006, pp. 293--296.

\end{thebibliography}

\newpage

\vfill

\end{document}